# A Generative deep learning approach for shape recognition of arbitrary objects from phaseless acoustic scattering data

W. W. Ahmed[1], M. Farhat[1], P.-Y. Chen[2], X. Zhang[3†], and Y. Wu[1*]

[1]*Division of Computer, Electrical and Mathematical Sciences and Engineering, King Abdullah University of Science and Technology (KAUST), Thuwal, 23955-6900, Saudi Arabia*

[2]*Department of Electrical and Computer Engineering, University of Illinois at Chicago, Chicago, IL60607, United States of America*

[3]*Department of Computer Science and Engineering, University of Notre Dame, Notre Dame, IN 46556, United States of America*

Emails: [*]ying.wu@kaust.edu.sa, [†]xzhang33@nd.edu

**Abstract:** We propose and demonstrate a generative deep learning approach for the shape recognition of an arbitrary object from its acoustic scattering properties. The strategy exploits deep neural networks to learn the mapping between the latent space of a two-dimensional acoustic object and the far-field scattering amplitudes. A neural network is designed as an Adversarial autoencoder and trained via unsupervised learning to determine the latent space of the acoustic object. Important structural features of the object are embedded in lower-dimensional latent space which supports the modeling of a shape generator and accelerates the learning in the inverse design process. The proposed inverse design uses the variational inference approach with encoder and decoder-like architecture where the decoder is composed of two pre-trained neural networks: the generator and the forward model. The data-driven framework finds an accurate solution to the ill-posed inverse scattering problem, where non-unique solution space is overcome by the multi-frequency phaseless far-field patterns. This inverse method is a powerful design tool that doesn't require complex analytical calculation and opens up new avenues for practical realization, automatic recognition of arbitrary shaped submarines or large fish, and other underwater applications.

## Introduction:

Wave propagation is widely used to solve the inverse scattering problem[1] (e.g., retrieving the shape of arbitrary objects from scattering data) in a variety of real-life engineering applications, including remote sensing[2], medical imaging[3], tomography[4], underwater robotics[5], geophysical prospecting[6], and detection of defects in oil and gas industry[7]. In the past decade, underwater activities such as biological research, target recognition, exploration of seabed resources, and monitoring of the underwater environment have increased the demand for subsea inspections. Compared to other waves, such as electromagnetic waves and visible light, sound waves have their own advantage in detecting targets in the long range because of their better transmissive characteristics. Several sonar devices have been developed to detect submarine objects using sound waves[8]. Sonar uses transmitter and sensor elements to send and receive acoustic waves. The waves are reflected by objects on the seabed and detected by the sensors; these measurements over time can be used to reconstruct sonar images with delay sum method to acquire underwater information. Despite of their success in detecting certain underwater objects, this method, however, does not adapt well to the complex and constantly changing environment. In this way, the recognition of underwater targets is still primarily dependent on decision of trained sonar operators[9], which can be highly imprecise due to the need of continuous manual monitoring. As a result, an automatic and reliable recognition approach must be developed to take over human tasks. In addition, scattering information is complex-valued radiated data, whose intensity and phase may not always be accurately measured. Especially it is difficult to retrieve the phase information from the scattered field data, and only the intensity information is available. Several analytical and numerical methods have been proposed to reconstruct the scatterer's shape form the scattering data, such as regularization[10], factorization[11], linear sampling method[12], and many others[13-15], to name a few. Yet, these computation-driven and brute force optimization methods are time consuming and require tedious hit-trail efforts in order to achieve the desired performance.

Recent advances in machine learning (ML) have made deep learning approaches an efficient way to solve forward and inverse scattering problems[16,17]. Deep learning networks can approximate the true solution to on-demand inverse designs, attributed to their ability to learn nonlinear mappings in datasets. Several studies have focused on discriminative and generative design methods and show excellent performance beyond the human capability[18-20]. Meng *et al.* presented a linear sampling method (LSM) with neural networks to reconstruct the shape of obstacles with acoustic far-field data[21]. LSM relies on selecting a contour line to obtain the shape information of an object. This solution is still somewhat imprecise, and requires further improvement. Fan et al. used convolutional neural networks (CNN) to determine the forward scattering properties for 2D convex prism geometries, but did not address the inverse problem[22]. Using neural networks, a compressing sampling matching pursuit (CoSaMP) method is developed to reconstruct the shape of arbitrary objects, but it only works for scatterers with low contrast relative to the surrounding environment[23]. In addition, probability-density-based deep learning approaches have also been developed to solve the inverse problem of the acoustic metasurfaces[24] and acoustic cloaks[25,26]. In both forward and inverse designs, the proposed solutions work well for simple geometries, but their performance decreases as the degree of freedom in the design space increases, making the scaling of complex models difficult. In this scenario, the generative models are used to reduce the dimensionality of the design space and efficiently learn the relations between design parameters and system responses[27,28]. For an arbitrary configuration, the design space is significantly enlarged and the inverse design process becomes extremely challenging due to one-to-many structure-property mapping, allowing diverse predictions. Zhang *et al.* presented an inverse design method for random metasurfaces, but the method is unable to entirely eliminate non-unique space because patterns and electromagnetic responses do not have one-to-one correspondence[29]. Lai *et al.* studied the inverse scattering problem that determines the configuration of 2D rigid cylinders for given total scattering cross sections using Wasserstein generative adversarial networks but had limited success in reducing the non-unique predictions[30]. In some situations, these disparate predictions may make the implementation easier, however, non-unique solutions drastically decrease the performance of the shape recognition algorithm in the far-field scattering problem. Deep neural networks have been designed and tested to predict the far-field scattering from 2D and 3D arbitrary objects[31,32], but the inverse scattering problem remains to be investigated. This study aims at solving the inverse scattering problem with deep learning that '*uniquely*' determines the shape of the object from solely its phaseless far-field information. To address the intrinsic one-to-many mapping problem[33], we feed multifrequency far-field data into the training process, thereby eliminating the non-unique solution space in final predictions. The inverse design procedure is as follows: Firstly, we encode the structural properties of the arbitrary shaped object in lower-dimensional latent vector, $z$, by using the adversarial autoencoder whose decoder part acts a generator, $G$, in the inverse design strategy [see Fig. 1(b)]. The designed adversarial autoencoder imposes the condition of gaussian distribution on the latent space to each randomly generated geometry and expedite the learning process from the shape of object to a given far-field profile. Secondly, we employ a forward neural network (FNN) that acts as a physics predictor to evaluate the inverse design process [see Fig. 1(c)]. In the final step, we probabilistically train an inverse neural network (INN) followed by the pretrained generator and forward simulator. After all networks are trained, far-field patterns are fed as an input to determine the latent space which predicts the generated design to the corresponding geometry [see Fig.1(c)]. This study presents a powerful method for modeling inverse far-field scattering that allows fast and accurate predictions of random shape of objects from scattering data for different applications.

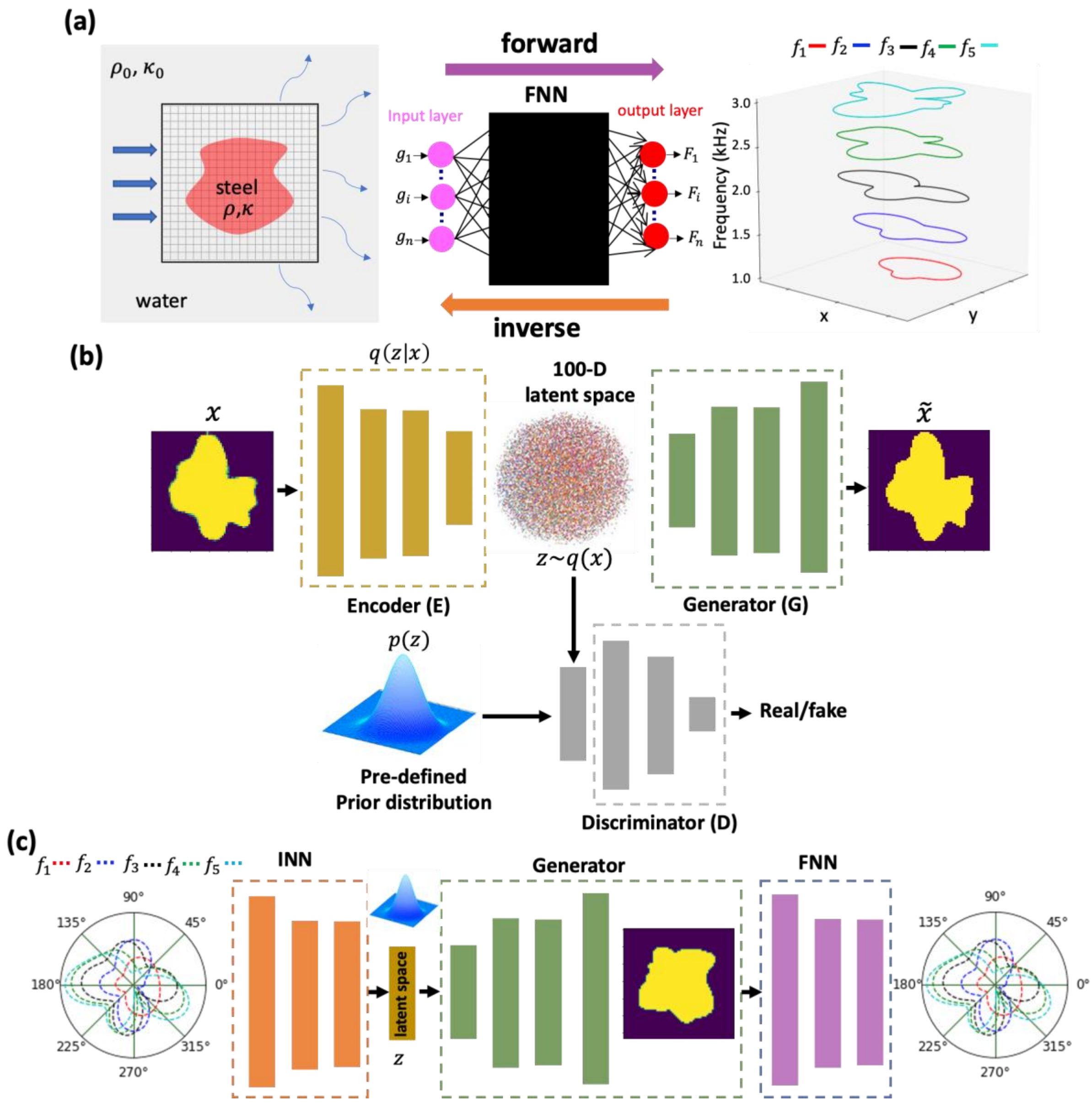

**Figure 1 | Machine learning assisted design process for acoustic far-field scattering problem.** (a) Forward and inverse mapping between the shape of object and far multi-frequency phaseless far-field amplitudes. (b) Architecture of Adversarial Autoencoder (AAE) to learn the latent space of arbitrary object with pre-defined model distribution. The typical autoencoder reconstructs the image, $x$, from latent space, $z$. A second discriminative network is trained to predict whether a sample originates from hidden data of an autoencoder or a user-specified distribution. (c) Inverse design approach where the pretrained Generator ($G$) and forward neural network (FNN) are cascaded behind the inverse neural network (INN). The generator $G$ extracted from the AAE and the trained FNN remain fixed and the INN is then trained stochastically to learn the mapping between the far-field patterns and the latent distribution of the object. After training, the latent space, $z$, is the design layer which is passed into

$G$ to predict the shape of the object for the given far-field information. In our study, we consider far-field amplitudes at five different frequencies in low frequency to unique identification of the object.

**Results and Discussion:**

We consider a 2D arbitrary-shaped scatterer made of steel immersed in water. Our goal is to build a deep learning model that can instantaneously provide information about the shape of the scatterer when illuminated by an acoustic plane wave. The scatterer is discretized into a pixel-based binary image where we assign the pixel value zero (for water) or one (for steel). The neural network is fed a binary image of the scatterer's shape as the input. The output of the network represents the directivity of the radiation pattern in the angular range of $0^{\circ}$ to $360^{\circ}$. A neural network can be designed to learn the isomorphic relation between the input and the output of the considered acoustic system as schematically described in Fig. 1(a).

To train the network, we need to prepare training samples. We randomly generate 20000 geometries to ensure diversity in training data sets. The far-field radiation properties of these random geometries are simulated with COMSOL Multiphysics[34] at five different frequencies $f_1 = 1$ kHz, $f_2 = 1.5$ kHz, $f_3 = 2$ kHz, $f_4 = 2.5$ kHz and $f_5 = 3$ kHz. The far-field data set contains 87 discrete points equally distributed in the full angular range ( $0^{\circ}$ to $360^{\circ}$). Each training sample consists of a pair of model input and the expected corresponding output: the 2D geometry and the far-field radiation pattern. The left panel in Fig. 1(a) illustrates an example of the structure with the red area representing steel and the corresponding radiation characteristics. The matrix of the structure is $64 \times 64$. The full data set contains 16, 000 (80%) training samples, 2000 (10%) validation samples, and the remaining 2000 (10%) testing samples. The validation data set monitors the overfitting during training and helps to tune the hyper-parameters of the network. The testing set evaluates the performance of trained network.

The proposed deep-learning model involves the training of adversarial autoencoder (AAE), forward neural network (FNN) and inverse neural network (INN). The AAE deals with the dimensionality reduction problem and learns the latent distribution of the arbitrary-shaped object to generate the geometric patterns. The FNN deals with the regression problem between the 2D object with the matrix of dimension 64 × 64 (equivalent to 1 × 4096-dimensional vector) and the multifrequency far-field scattering amplitudes with the dimension of 1 × 435. The INN relies on the generator of AAE and FNN to predict the shape of an arbitrary structure of dimension 64 × 64 for the given multifrequency far-field amplitudes of dimension 1 × 435. We will discuss the training of these three networks in the following sections.

### A. Adversarial Autoencoder (AAE)

We employ an AAE[35] to learn the distribution of random geometries whose training is based on traditional reconstruction and an adversarial regularization. The architecture of AAE is composed of three fully-connected coupled neural networks: encoder, generator, and discriminator, as shown in Fig. 1(b). In the AAE, the encoder transforms a given input geometry into a compressed continuous design space (a latent space); the generator reconstructs the real space geometry from a given latent space; and the discriminator forces the latent space to follow specified prior distribution by an adversarial manner.

The training process of AAE includes both adversarial learning of generative adversarial network (GANs) [36] and latent distribution learning of variational autoencoders (VAEs)[37]. In this sense, the AAE architecture incorporates the properties of VAEs sand GANs. Compared to other deep generative networks, such as VAEs and GANs, AAE networks possess several advantages. The AAE is more flexible than VAEs because it offers the freedom to control the latent space distribution without imposing any restrictions. It also produces a dense (continuous) latent space that can be used to generate diverse designs. Additionally, AAE networks are easier to train than GANs because they directly adversarial learning to a latent representation, rather than the geometry like GANs. Training the AAE network takes the generated datasets of arbitrary 2D geometries and maps them in a latent space, where the generator can reconstruct the 2D geometries. Meanwhile a random sample from the latent space is fed into the discriminator to be distinguished from a sample from a given prior distribution. The AAE is trained with binary cross-entropy loss function defined as: $L_{BCE} = -\frac{1}{N}\sum_{k=1}^{n} y_k \log{(\hat{y}_k)} + (1 - y_k)\log{(1 - \hat{y}_k)}$ where $N$ is the size of output, $y_k$ is the true probability value and $\hat{y}_k$ is the predicted probability value. When the AAE network is trained, the generator will be able to generate designs in real-space based on the latent vector. The trained generator is extracted from AEE and then used in the inverse design process. The training and prediction results for AEE network are shown in Fig. 2. Figures 2(a) and 2(b) depict the loss of the generator and discriminator network during the training process. The loss function of the generator smoothly converges as the number of epochs increase; however, the loss of the discriminator fluctuates due to adversarial learning. In order to evaluate the performance of trained AAE, we use the binary cross-entropy error (BCE) and structure similarity index measure (SSIM) as shown in Figs. 2(c) and 2(d), respectively. The details about the SSIM are provided in Supplementary Information (SI)[38]. The mean average values of BCE and SSIM are found to be 0.02 and 0.94, respectively. Two representative examples of the reconstructed geometries are shown in Figs. 2(e) and 2(f). The results clearly indicate the excellent prediction performance of the trained network.

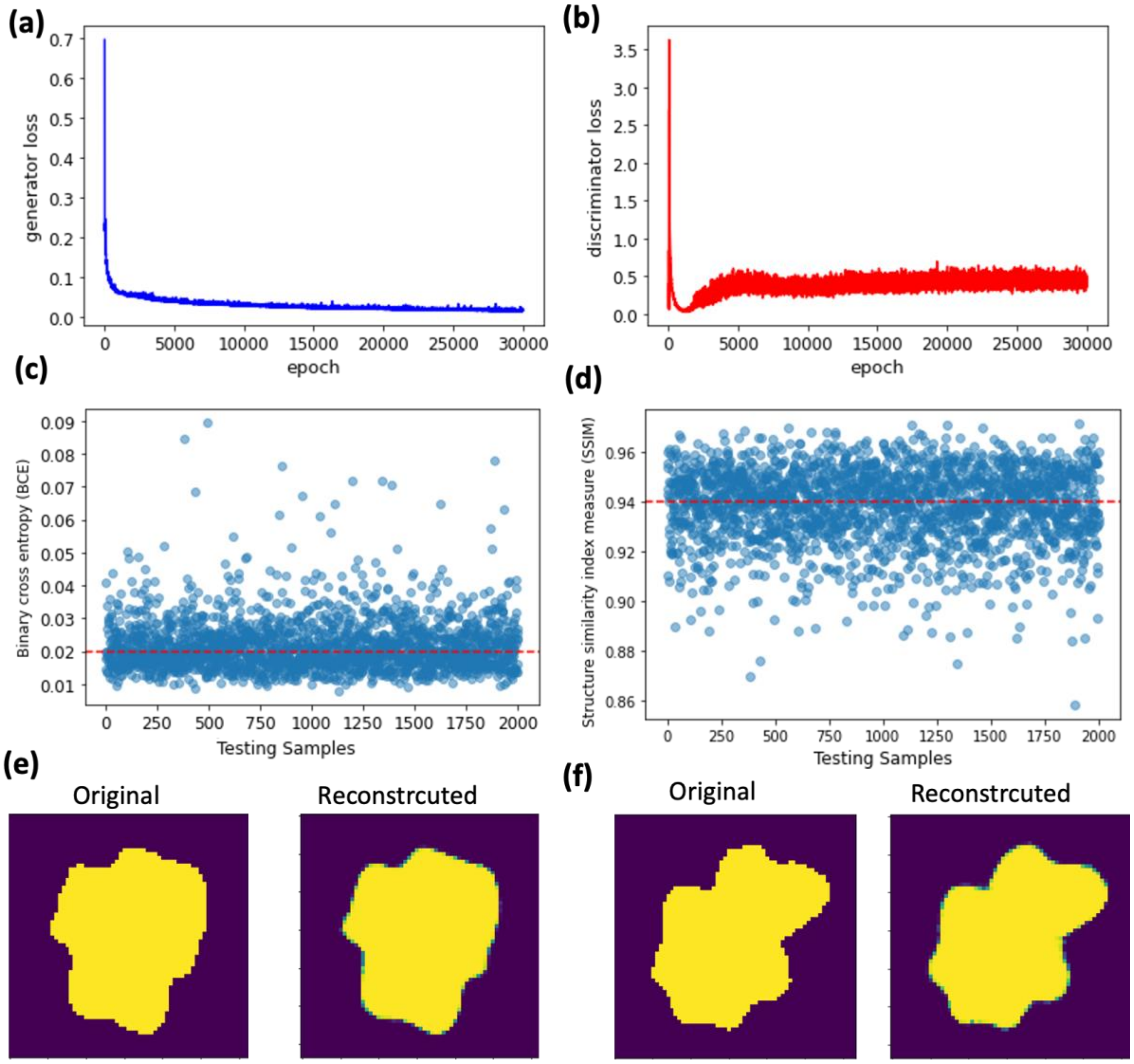

**Figure 2 | Designed AAE to learn the latent space of arbitrary binary structures.** (a) Generator binary cross-entropy loss function. (b) Discriminator binary cross-entropy loss function. (c) Binary cross-entropy error. (d) Structure similarity index measure (SSIM) on the test dataset where the red dashed line shows the average error. (e-f) Representative examples of original and reconstructed shapes of arbitrary objects from trained AAE network.

## B. Forward Neural Network (FNN)

Forward modelling consists of predicting the response of the given physical system. Different from the traditional approach, the data-driven approaches predict the acoustic response of a given structure without explicitly solving acoustic wave equations. To be more specific, we design a deep neural network that solves the regression problem deterministically by determining the far-field radiation patterns for a given arbitrary structure. The network learns the complex relation between arbitrary 2D binary structures and their associated multifrequency far-field radiation patterns. These patterns are invariant under translation of the object so each binary structure is flattened to feed the fully connected network as an input. For the training process, mean squared error (MSE), is used as the loss function that can be expressed as $\mathcal{L} = \frac{1}{N}\sum_i \left(F_i - \hat{F}_i\right)^2$ where $F_i$ and $\hat{F}_i$ are the ground truth and the predicted far-field pattern, respectively. The architecture of FNN contains $4096 - 1000 - 1000 - 800 - 800 - 800 - 800 - 600 - 600 - 600 - 435$ nodes and the details of training process and network hyper parameters are provided in SI[37]. The learning behavior as function of different epochs is shown in Fig. 3(a). In order to verify the validity and accuracy of the designed network, we use the test dataset to calculate the relative absolute error as follows: $e = \sum_i \left|F_i - \hat{F}_i\right|/F_i$ where $F_i$ and $\hat{F}_i$ correspond to target and predicted far-field response. Fig. 3 (b) illustrates the distribution of error and corresponding mean relative error of the test set predictions is below 4%, as indicated by the dashed vertical red line. A typical example shows a good match between the COMSOL Multiphysics simulation and the FNN prediction results [see Fig. 3 (c)]. An accurate training of FNN is crucial because the following inverse design approach uses the trained FN as an essential component for accurate and instant prediction of far-field patterns for the given object.

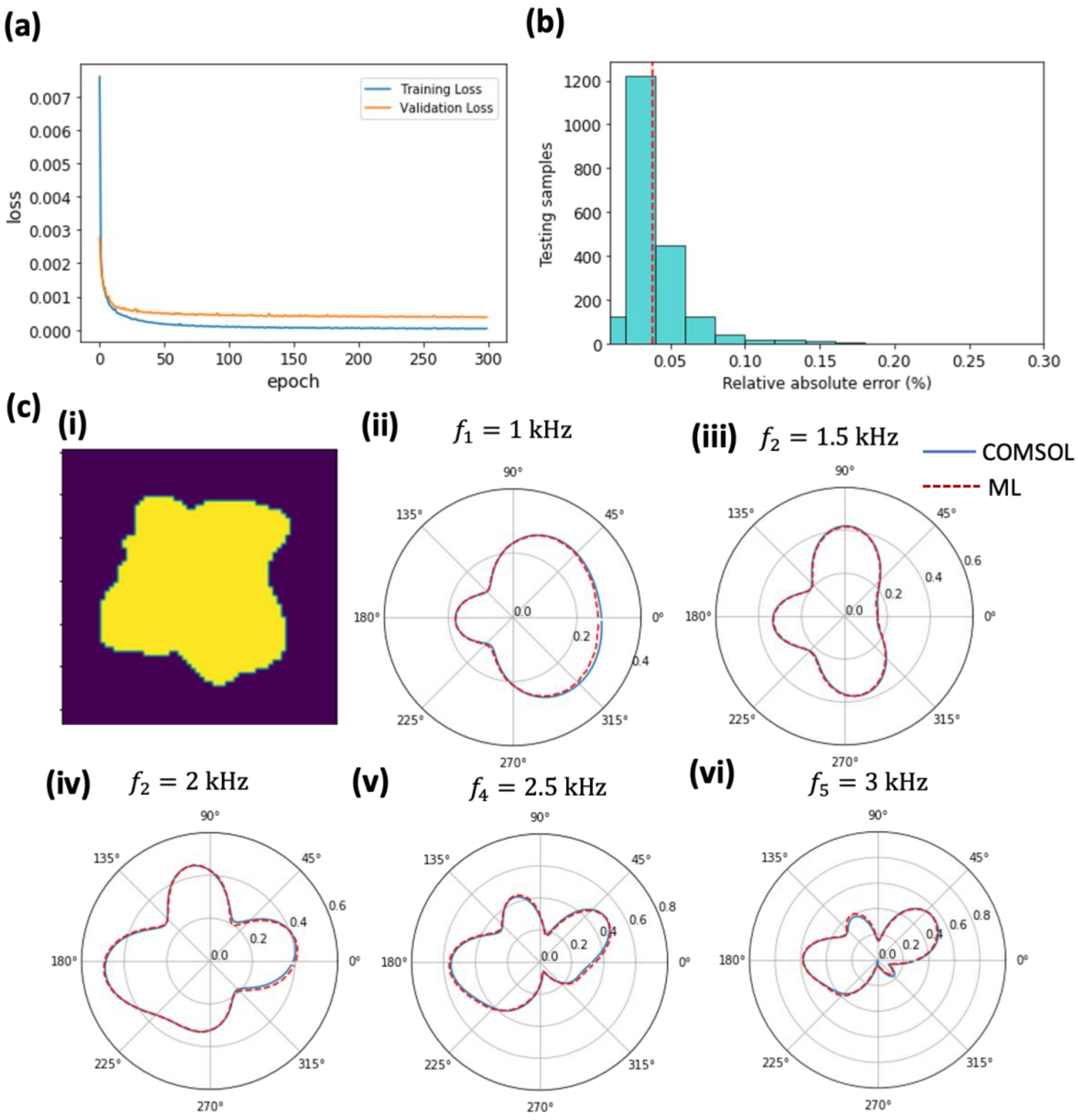

**Figure 3 | Designed FNN for arbitrary-shaped scatterer to multifrequency far-field mapping.** (a) Evolution of MSE loss function. (b) Absolute prediction error on the test dataset where the red dashed line shows the average error. (c) Representative examples of predicted multifrequency far field amplitudes for the trained networks: (i) an arbitrary binary structure, (ii-vi) far-field profiles at different frequencies where solid blue and dashed red curves indicate the target and predicted results.

### C. Inverse Neural Network (INN)

The next step extends deep learning to solve the inverse design problem, which is essentially the inverse process of the forward modelling. While the forward modeling deals with a one-to-one mapping between a physical system and the resultant response, the inverse design is facing the difficulty induced by the non-unique solution spaces. Therefore, a single discriminative network is not able to learn the complex relation in the inverse design. To circumvent this problem, auxiliary training approaches, generative models, and optimization strategies are combined in the inverse design process[38-44]. The most common method to overcome the non-uniqueness issue is exploiting a tandem network that incorporates the forward modelling network into the inverse design DNN architecture. In conventional tandem networks, direct learning of the pixel-binary structure is not trivial due to the high degree of freedom in parameter design. In addition, the conventional tandem networks do not guarantee the complete elimination of non-unique solution space. Here, we propose a new solution of generative tandem network which learns the mapping between the far-field patterns and the latent distribution of binary structure by incorporating an additional pretrained network, i.e., the shape generator extracted from trained AAE, into our inverse design strategy. The INN translates the input far-field patterns into the latent representation $z$, sampled from the Gaussian prior distribution parametrized by mean $\mu$ and standard deviation $\sigma$, to approximate the latent space of an arbitrary geometry as illustrated in Fig.1(c). The loss function for the INN is constructed by mean absolute loss for reconstruction of far-field patterns, $\mathcal{L}_{MAE}$ and Kullback-Leibler (KL) divergence $\mathcal{L}_{KL}$ between latent space distribution and prior Gaussian distribution, $\mathcal{N}(0,1)$. To be specific, the training of INN is to minimize the following loss function expressed as:

$$\mathcal{L} = \frac{1}{N}\sum_{n=1}^{N}\left(\mathcal{L}_{MAE}^{(n)} + \alpha\mathcal{L}_{KL}^{(n)}\right), \tag{1}$$

where $N$ is the total number of training samples and $\alpha$ is the relative weight between deterministic and generative learning. The details about the stochastic tuning of the model are provided in the method section.

The INN architecture is composed of eight layers with each layer having $435 - 800 - 800 - 500 - 500 - 500 - 400 - 100$ nodes. The intermediate latent space layer acts as a design layer, which is then passed into the generator and forward modeling part to calculate the corresponding far-field pattern. The learning curve is shown in Fig. 4(a). The rapidly decreasing behavior of learning curve demonstrates that training is highly effective. To evaluate the performance of the INN, we calculate the prediction errors on inversely generated far-field patterns, i.e., relative absolute error, and on the predicted binary structure, i.e., binary cross entropy, and structure similarity index measure. The mean percent error of inverse-designed far-field patterns is around 3% [see Fig. 4(b)], while the average BCE error and SSIM are 0.077 and 0.85, respectively [see Figs. 4(c) ad 4(d)].

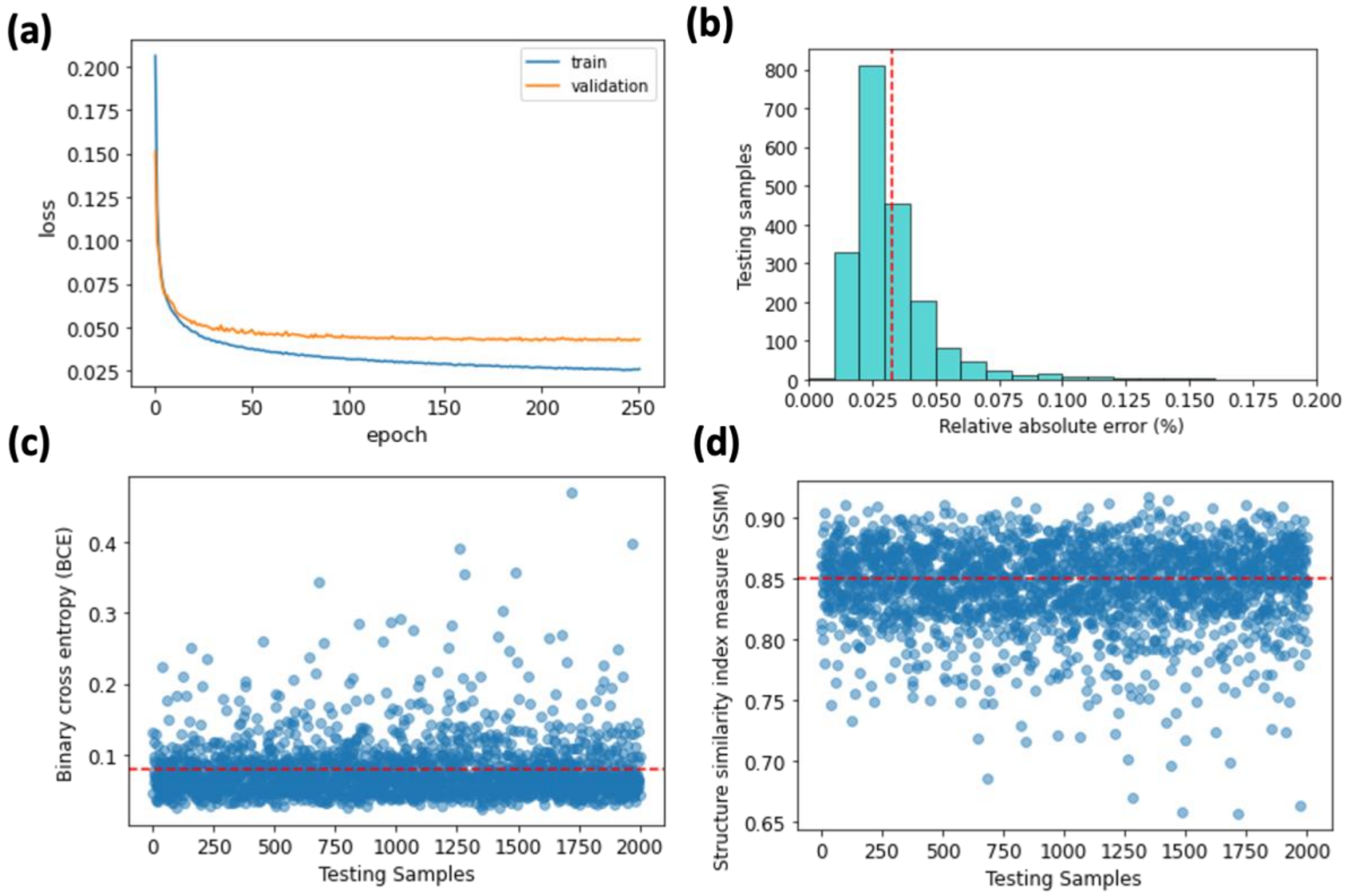

**Figure 4 | Designed INN for shape recognition from given multifrequency far-field patterns.** (a) Evolution of the loss function of the generative inverse model. (b) Absolute prediction error for reconstructed far-field patterns. (c) Binary cross-entropy error. (d) Structure similarity index on predicted binary structures. The red dashed line indicates the average error on test datasets in (b-d).

Figure 5 illustrates two examples to showcase how the trained inverse network accurately determines the shape of an object for the given multi-frequency far-field patterns. The predicted far-field patterns from the generated shapes are provided to compare with the given far-field patterns. As we can see, inversely designed objects (generated) for the given far-field pattens match well with the target objects, which confirms the efficacy of our method. The small variations around the edges barely affect on the corresponding far-field information [see Figs. ab (iii)-(vii)].

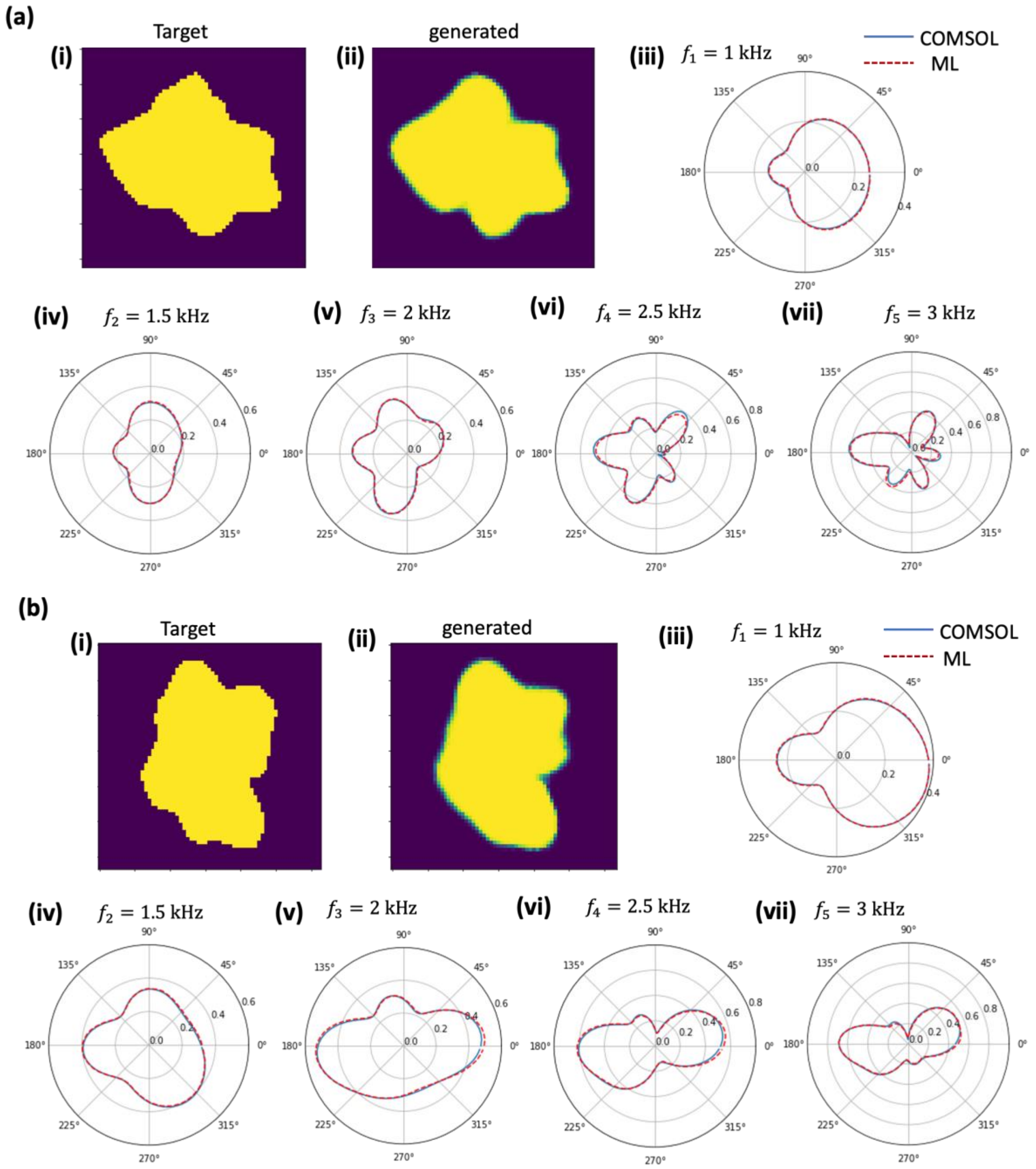

**Figure 5 | Examples of generated shapes from given multifrequency far-field patterns using INN.** (i) Target structure. (ii) Inversely generated structure from far-field patterns. (iii)-(vii) Comparison of the simulated (solid black curve) and predicted far-field pattern (dashed red curves) from the generated geometries in (a, b).

Next, we study the effect of feeding the multifrequency data for *'unique'* shape recognition. We investigate the performance of the INN by starting with a single frequency far-field data, 1 kHz, and then continuously increasing the number of frequencies in discrete steps of 0.5 kHz. For brevity, this study limits the INN results to five frequencies. The INN results are shown in Fig. 6. We observe that the performance of the network to uniquely identify a structure is improved as the number of frequencies considered for the far-field data increases. Figure 6(a) clearly depicts that the binary cross-entropy error used to identify an arbitrary shape sharply decreases when the second frequency is added, and then slowly saturates as more frequencies are taken into account. A similar trend is observed with SSIM, where more frequencies make the SSIM approaching towards unity [see Fig. 6(b)]. Thus, we can conclude that adding more frequencies in network training, the network gives better results. We found five frequencies are sufficient for an optimal performance (further increasing the frequencies number does not noticeably improve the performance). The detailed results for the trained FNN and INN for multi-frequency data are provided in SI.

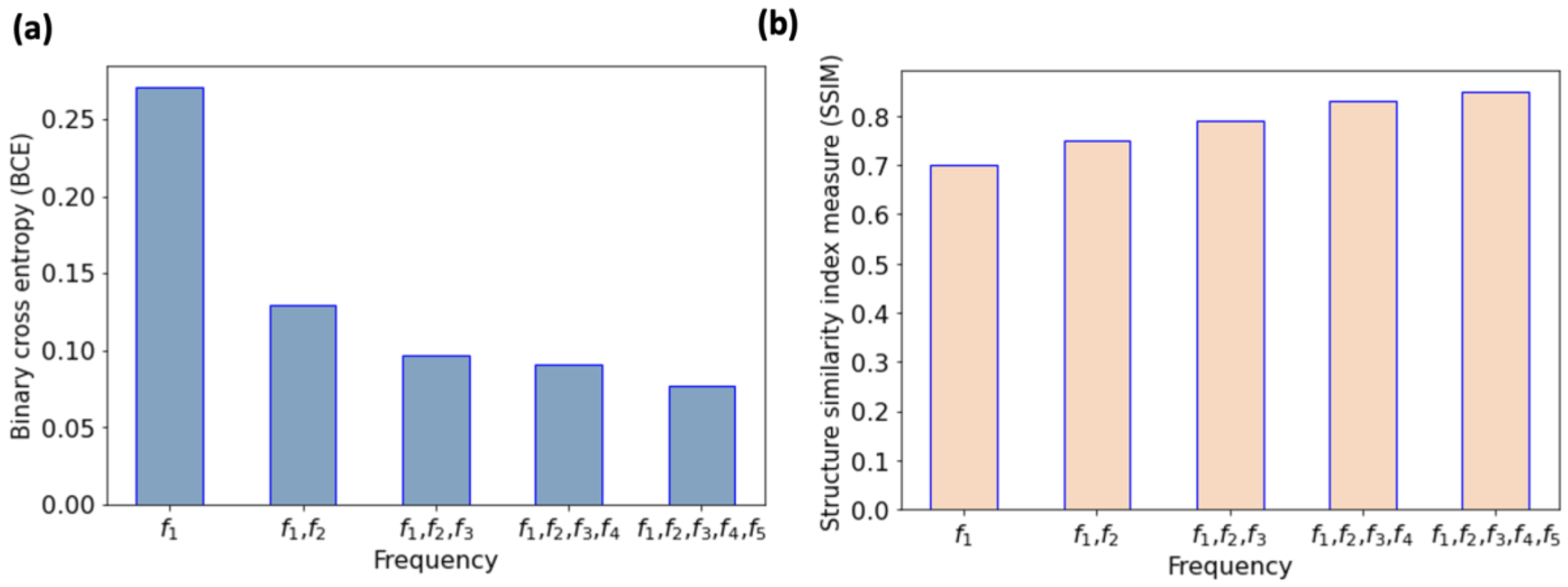

**Figure 6 | Comparison of INN performance with multifrequency far-field data.** (a) Frequency dependent Binary cross entropy error and (b) structure similarity index measure (SSIM). BCE error decreases and SSIM

increases with feeding multifrequency far-field data to the neural network where $f_1 = 1$ kHz, $f_2 = 1.5$ kHz, $f_3 = 2$ kHz, $f_4 = 2.5$ kHz and $f_5 = 3$ kHz. The accuracy of INN to recognize the shape is significantly improved with multifrequency data.

In reality, it may not be feasible to measure multi-frequency far-field data over the entire angular range of $0^\circ$ to $360^\circ$. Therefore, analyzing the performance of the network with partial angular field data is crucial. For the sake of demonstration, we provide far-field data during training in a reduced angular range of $0^\circ$ to $180^\circ$ [see inset in Fig.7(a)], and the results of training the network are shown in Fig. 7. Although the average error of prediction is slightly compromised because of reduced information used for learning in the training, the proposed method can still accurately identify the key features of those arbitrary objects. The detailed results are shown in SI.

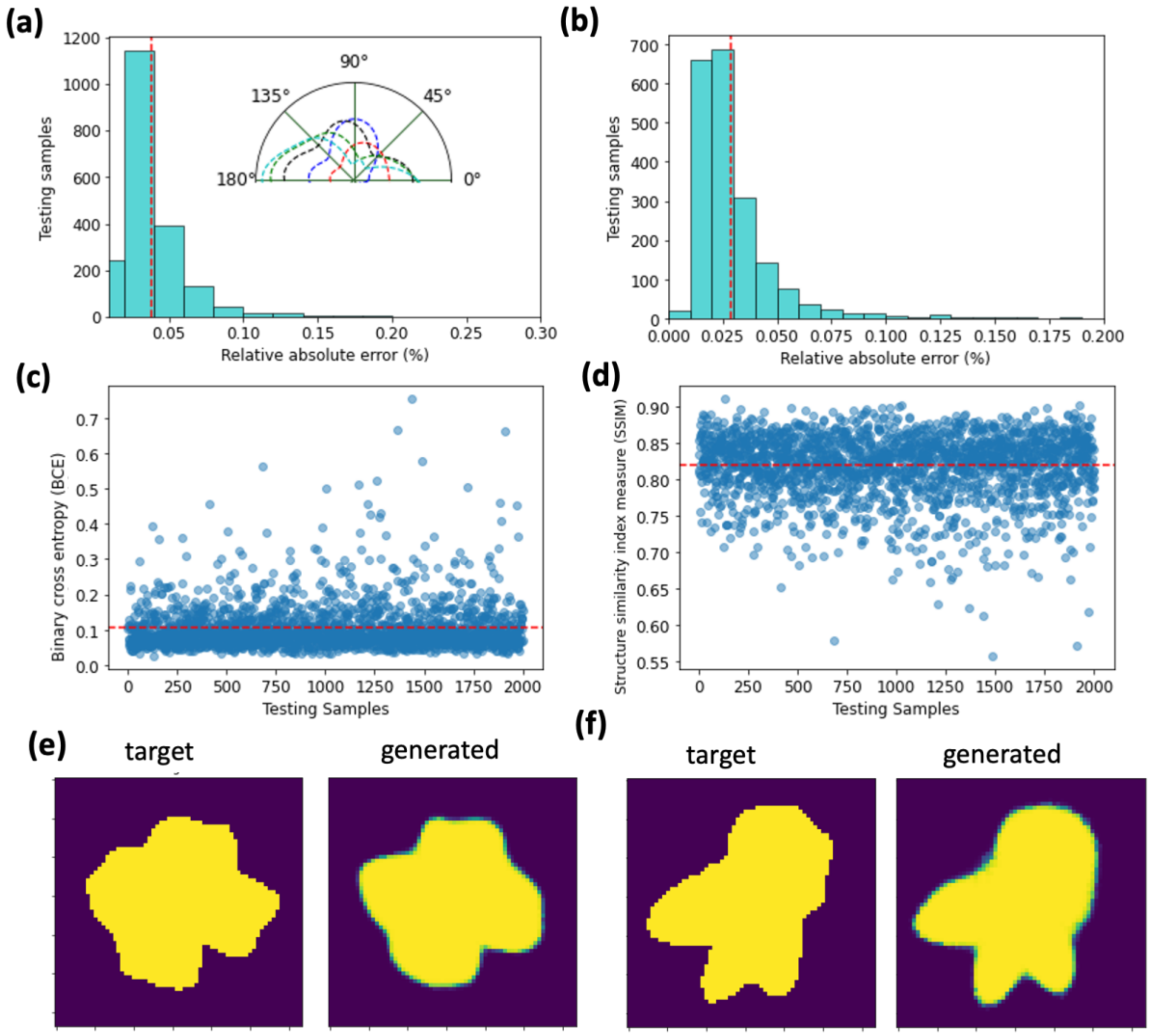

**Figure 7 | FNN and INN results for multifrequency half-plane far-field data.** The distribution of absolute relative error to predict the far-field through trained FNN (a) and INN (b). The inset in (a) shows the half-plan multi-frequency far-field patterns for an arbitrary-shaped object. The binary cross-entropy error and structure similarity index measure of INN to recognize the shape are shown in (c) and (d), respectively. (e-f) Examples of generated geometries for the given far-field patterns. [See the SI for details].

## Conclusion

The inverse scattering problem of retrieving the shape from the far-field information is highly nonlinear and extremely challenging to solve with conventional approaches due to non-uniqueness issues. Here, we propose a generative deep learning approach as a practical design tool, to uniquely determine the shape of an arbitrary object with multifrequency phaseless data. We exploit generative adversarial learning to encode the true features of objects into the latent space through adversarial autoencoder and further integrate its generator into the inverse design process to create random shapes. The forward network is designed to learn the relation between a given structure and the corresponding far-field profile for instant and accurate predictions. The inverse design strategy is based upon a generative encoder-decoder-like architecture, where the encoder (i.e., inverse network) is trained while fixing the decoder consisting of the pretrained forward network and generator. We study the influence of feeding multifrequency far-field data on shape recognition and show that multifrequency data rules out the non-unique solution spaces in the inverse architecture. In addition, we demonstrate that half angular far-field data (i.e., $0^{\circ}$ to $180^{\circ}$) is still capable of uniquely determining the shape of the arbitrary object. The designed network instantly predicts the shape of the object for the given far-field information with unprecedented speed, reducing the design time by four orders of magnitude over traditional methods. The proposed approach has potential applications for automatic detection of underwater objects such as submarines, fish species and monitoring of the others underwater activities. Our approach is generic and readily extended to design complex and random geometries of three-dimensional space structures and other physical systems, such as optics, plasmonic, to name a few.

## Methods:

### Adversarial Learning Method:

In AAE, the encoder takes a real image $x \in R^d$ as an input, compresses it into a $p$-dimensional latent space $z \in R^p$ (where $p \ll d$), and then the generator reconstructs the image $\tilde{x} \in R^d$. Let $q(z)$ be the posterior distribution of latent space, $z$, in autoencoder and $p(z)$ be user-specified prior distribution, which is assumed to be Gaussian distribution. The encoder generates the latent space from the posterior distribution: $z \sim p(z)$ and discriminator $D$ distinguishes $z$ to be a real from the prior distribution $p(z)$ or generated by the encoder as follows:

$$D(z) = \begin{cases} 1 & \text{if } z \sim p(z) \\ 0 & \text{if } z \sim q(z). \end{cases} \tag{2}$$

AEE attempts to generate $z$ analogous to the real latent space from the prior distribution via adversarial learning.

**Variational Inference Method:**

The inverse network exploits the variational inference method to learn the distribution of far-field response and its mapping with the latent space, $z$, of the arbitary object. The reconstruction loss is the MAE between the input far-field patterns and the reconstructed patterns by the INN. The learned distribution $q(z|x)$ equivalent to the predefined distribution $p(z)$ is ensured through KL divergence term in loss function expressed as

$$\mathcal{L}^{(n)} = \mathcal{L}_{MAE}^{(n)} + \alpha \sum_{p} \mathcal{L}_{KL}^{(n)}\left(q_p(z|x) \parallel p(z)\right), \tag{3}$$

where $n$ represents the $nth$ dataset, $p$ is the dimensionality of the latent space $z$, $\alpha$ is the relative weight between the deterministic and stochastic learning. In our study, we tunned $\alpha = 10^{-5}$ for accurate prediction of the shape of object via INN. The KL divergence between two Gaussian distributions can be defined as

$$\mathcal{L}_{KL}^{(n)} = -\frac{1}{2}\sum_{p} 1 + \sigma_p^2 - \mu_k^2 - lo\,g\left(\sigma_p\right) , \tag{4}$$

where $\mu$ and $\sigma$ are mean and standard deviation of the generated latent space distribution, respectively. The reparameterization trick helps to determine $z = \mu_p + \sigma_p \varepsilon$, where $\varepsilon$ is a sample from prior Gaussian distribution, during the optimization process. The reconstruction term is expressed as the MSE between the input coding matrix and the reconstructed coding matrix output by the decoder.

**Numerical modeling:**

The finite element method is used to perform the full wave simulations with COMSOL Multiphysics. We assume water as the background medium in our simulation setup with mass density $\rho = 1000\ \mathrm{kg/m^3}$, bulk modulus $\kappa = 2.91$ GPa. The arbitrary object is made of steel with mass density $\rho = 7850\ \mathrm{kg/m^3}$, bulk modulus $\kappa = 201$ GPa. The steel object in water produces shear waves so we model the steel with solid mechanics modules and background medium with acoustic module. The integration of acoustic module and solid mechanics model is done with Multiphysics module for full wave modelling. The arbitrary object is excited with an acoustic plane wave of unit amplitude to determine the far-field response at different operating frequencies. [See SI for details on the derivation and governing equations.]

## Acknowledgements

The work described in here is supported by King Abdullah University of Science and Technology (KAUST) Artificial Intelligence Initiative Fund and KAUST Baseline Research Fund No. BAS/1/1626-01-01.

## Supplementary information

### 1. Elastic wave interaction with solid objects modeling

Let us consider the elastic cylinder shown in Fig. S1, with radius $a$, density $\rho_1$, and speeds of sound $c_1$ and $c_2$, corresponding to pressure and shear waves, respectively.

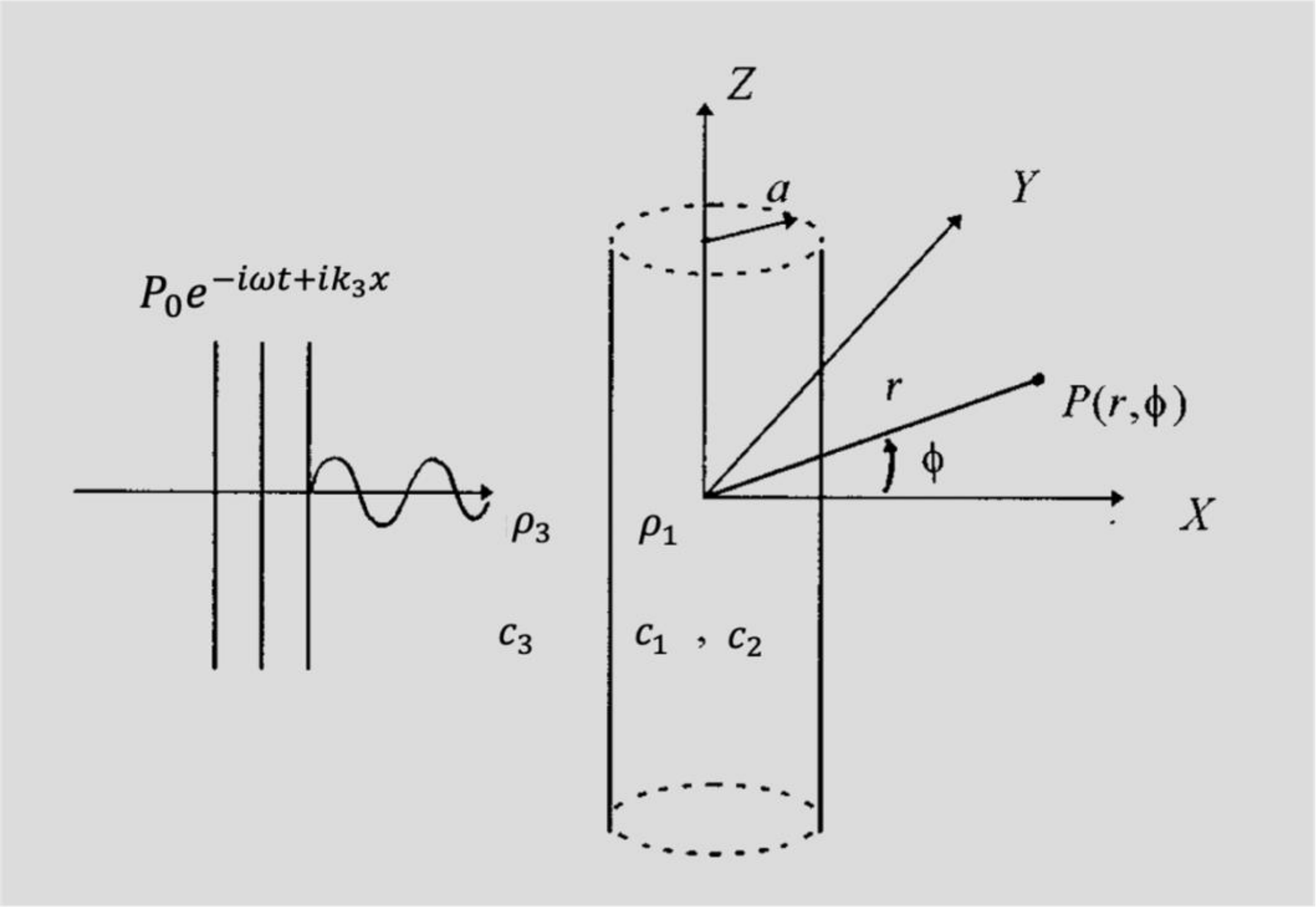

Fig. S1. Scheme and geometry of the acoustic wave scattering from an infinitely extended (in the $z$ −direction) elastic cylinder.

### 1.1 General governing equations

The general (linear) equation governing the propagation of elastic waves in solids (taking into account shear waves) is given by

$$(\lambda + \mu)\nabla(\varsigma) - \mu\nabla \times (2\boldsymbol{\Omega}) = \rho_1 \frac{\partial^2 \boldsymbol{u}}{\partial t^2}, \quad (1.1)$$

with $\varsigma = \nabla \cdot \boldsymbol{u}$ and $2\boldsymbol{\Omega} = \nabla \times \boldsymbol{u}$. From Eq. (1.1) we can derive the equations for $\varsigma$ and $2\boldsymbol{\Omega}$ as

$$\Delta\varsigma = \frac{\rho_1}{\lambda + 2\mu}\frac{\partial^2 \varsigma}{\partial t^2}, \quad (1.2)$$

and

$$\Delta(2\mathbf{\Omega}) = \frac{\rho_1}{\mu}\frac{\partial^2(2\mathbf{\Omega})}{\partial t^2}, \tag{1.3}$$

where the velocities can be defined as

$$c_1 = \sqrt{\frac{\lambda + 2\mu}{\rho_1}};\ c_2 = \sqrt{\frac{\mu}{\rho_1}}, \tag{1.4}$$

where $\lambda$ and $\mu$ are the Lamé parameters.

In order to solve Eq. (1.1) we need to rewrite the displacement vector as

$$\boldsymbol{u} = -\nabla\psi + \nabla \times \boldsymbol{A}, \tag{1.5}$$

to obtain

$$\Delta\psi = \frac{1}{c_1^2}\frac{\partial^2\psi}{\partial t^2} \text{ and } \Delta\boldsymbol{A} = \frac{1}{c_1^2}\frac{\partial^2\boldsymbol{A}}{\partial t^2} \tag{1.6}$$

To solve Eq. (5.5) in the case of cylindrical objects (Fig. 5.1) we can use the following ansatz

$$\psi = \sum_{n=0}^{\infty} a_n J_n(k_1 r)\cos n\theta, \tag{1.7}$$

$$A_z = \sum_{n=0}^{\infty} b_n J_n(k_2 r)\sin n\theta. \tag{1.8}$$

Equation (1.8) is shown to be valid as $\boldsymbol{A}$ can be shown to have components in the $z$-direction because of symmetry considerations. Also, only $\sin n\theta$ terms appear in Eq. (1.8) as the potential vector has to be anti-symmetrical about the axis $\theta = 0$, to ensure that the resulted displacement is symmetrical around $\theta = 0$.

Therefore, by combining Eq. (1.5) and Eqs. (1.7)-(1.8), we can get

$$u_r = \sum_{n=0}^{\infty}\left\{\frac{nb_n}{r}J_n(k_2 r) - a_n\frac{d}{dr}J_n(k_1 r)\right\}\cos n\theta, \tag{1.9}$$

$$u_\theta = \sum_{n=0}^{\infty} \left\{ \frac{na_n}{r} J_n(k_1 r) - b_n \frac{d}{dr} J_n(k_2 r) \right\} \sin n\theta. \quad (1.10)$$

On the other hand, the incident plane wave can be represented by

$$p_i = P_0 e^{-ik_3 x} = P_0 e^{-ik_3 r \cos\theta} = P_0 \sum_{n=0}^{\infty} \varepsilon_n (-i)^n J_n(k_3 r) \cos n\theta. \quad (1.11)$$

The radial component of displacement of this incident plane wave is thus (from now on we assume, without loss of generality, that $P_0 = 1$ to simplify the expressions)

$$u_{i,r} = \frac{1}{\rho_3 \omega^2} \frac{\partial p_i}{\partial r} = \frac{P_0}{\rho_3 \omega^2} \sum_{n=0}^{\infty} \varepsilon_n (-i)^n \frac{d}{dr} J_n(k_3 r) \cos n\theta, \quad (1.12)$$

and the scattered wave is

$$p_s = \sum_{n=0}^{\infty} c_n H_n^{(1)}(k_3 r) \cos n\theta, \quad (1.13)$$

and its associated radial component of displacement is

$$u_{s,r} \frac{1}{\rho_3 \omega^2} \sum_{n=0}^{\infty} c_n \frac{d}{dr} H_n^{(1)}(k_3 r) \cos n\theta, \quad (1.14)$$

where $c_n$ are the scattering coefficients, to be evaluated via the boundary conditions of the problem.

## 1.2 Boundary conditions and numerical simulations

In the case of an elastic cylinder in an acoustical media (such as water) the boundary conditions that shall be employed on the cylinder's boundaries are: (i) The pressure in the fluid must be equal to the normal component of stress at the solid interface; (ii) the radial component of displacement of the fluid has to be equal to the normal component of displacement of the solid interface; and (iii) the tangential components of shear stress must vanish at the interface of the solid. This means specifically

$$p_i + p_s = -[rr] \text{ at } r = a, \quad (1.15)$$

$$u_{i,r} + u_{s,r} = u_r \ \text{ at } \ r = a, \tag{1.16}$$

$$[r\theta] = [rz] = 0 \ \text{ at } \ r = a, \tag{1.17}$$

where the stress components in cylindrical coordinates are

$$[rr] = \lambda\varsigma + 2\mu\frac{\partial u_r}{\partial r} = 2\rho_1 c_2^2\left\{(\sigma\mathbb{I} - 2\sigma)\varsigma + \frac{\partial u_r}{\partial r}\right\}, \tag{1.18}$$

$$[r\theta] = \mu\left\{\frac{1}{r}\frac{\partial u_r}{\partial \theta} + r\frac{\partial}{\partial r}\left(\frac{u_\theta}{r}\right)\right\}, \tag{1.19}$$

$$[rz] = \mu\left\{\frac{\partial u_r}{\partial z} + \frac{\partial u_r}{\partial r}\right\}. \tag{1.20}$$

By applying the boundary conditions at the interface between the elastic cylinder and the surrounding fluid, we get the linear system

$$\begin{pmatrix} -x_1 J_n'(x_1) & nJ_n(x_2) & \frac{-x_3}{\rho_3\omega^2} H_n^{(1)'}(x_3) \\ x_1^2[-2\mu J_n''(x_1) + \lambda J_n(x_1)] & 2\mu n[x_2 J_n'(x_2) - J_n(x_2)] & a^2 H_n^{(1)}(x_3) \\ 2n[x_1 J_n'(x_1) - J_n(x_1)] & -n^2 J_n(x_2) + x_2 J_n'(x_2) - x_2^2 J_n''(x_2) & 0 \end{pmatrix}\begin{pmatrix} a_n \\ b_n \\ c_n \end{pmatrix} = \begin{pmatrix} \frac{x_3}{\rho_3\omega^2} J_n'(x_3) \\ -a^2 J_n(x_3) \\ 0 \end{pmatrix}. \tag{1.21}$$

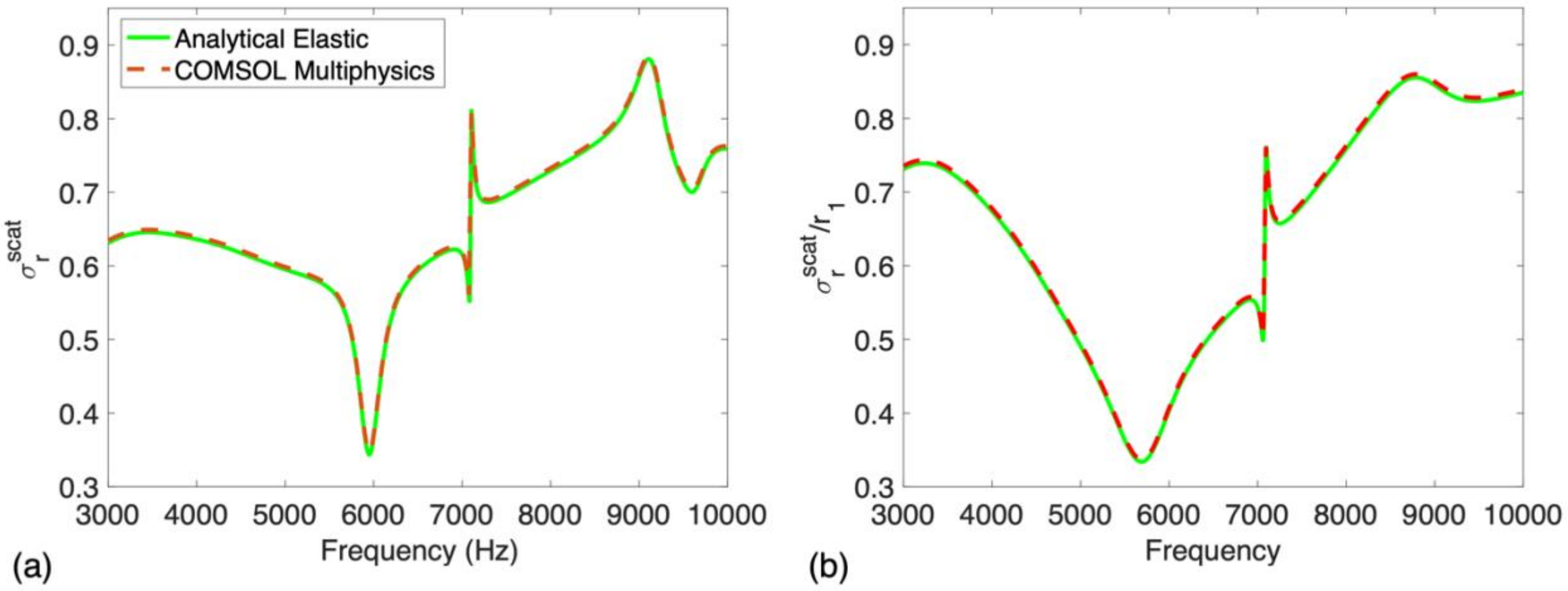

Fig. S2. Scattering cross-section from an elastic cylinder made of (a) steel and (b) Aluminum embedded inside water. The dashed (red) curve gives the COMSOL simulations whereas the green solid curves give the Mie results discussed in this section.

To obtain $c_n$ we make use of the Cramer's rule, i.e., $c_n = \frac{\det V_n}{\det M_n}$, where $V_n$ is the matrix obtained from $M_n$ by replacing its last column by the RHS of Eq. (1.21).

In order to verify the validity of the COMSOL Multiphysics elastic/acoustic modeling of scattering, we compute and give in Fig. S2 the SCS of a cylinder by using both COMSOL (red dashed lines) and the Mie formalism discussed in the previous paragraphs (green solid lines). These results match perfectly for both a steel cylinder [S2(a)] in water and an Aluminum cylinder [S2(b)] in water, validating thus the numerical modeling by COMSOL, that we use throughout this paper.

## 2. Structure Similarity Index Measure (SSIM

SSIM is a metric that quantify the *similarity* between two given images by extracting key characteristic from an image named Luminance, Contrast, and Structure. Luminance is measured by averaging over all the pixel value denoted by $\mu$ reads as: $\mu_x = \frac{1}{N}\sum_{i=0}^{N} x_i$. Contrast is measured by taking the standard deviation of all the pixel values denoted by σ reads as: $\sigma_x = \sqrt{\frac{1}{N-1}\sum_{i=0}^{N}(x_i - \mu_x)^2}$ and structure comparison is computed by $(x - \mu_x)/\sigma_x$. The luminance comparison function $l(x, y)$ is a function of $\mu_x$ and $\mu_y$ defined as:

$$l(x, y) = \frac{2\mu_x\mu_y + c_1}{\mu_x^2 + \mu_y^2 + c_1}$$

The contrast comparison function $c(x, y)$ is a function of $\sigma_x$ and $\sigma_x$ defined as:

$$c(x, y) = \frac{2\sigma_x\sigma_y + c_2}{\sigma_x^2 + \sigma_y^2 + c_2}$$

Structure comparison function $s(x, y)$is defined as:

$$s(x, y) = \frac{2\sigma_{xy} + c_3}{\sigma_x \sigma_y + c_3}$$

where $\sigma_{xy} = \sqrt{\frac{1}{N-1}\sum_{i=0}^{N}(x_i - \mu_x)^2(y_i - \mu_y)^2}$, and $c_1, c_2, c_3$ are constants to ensure stability when the denominator becomes 0.

And finally, luminance, contrast, and structure comparison functions are combined to determine the similarity index value given by,

$$\mathrm{SSIM}(x, y) = [l(x, y)]^{\alpha}[c(x, y)]^{\beta}[s(x, y)]^{\gamma}$$

where $\alpha > 0, \beta > 0, \gamma > 0$ denote the relative strength of each of the metrics. To simplify the expression, we assume, $\alpha = \beta = \gamma = 1$ and $c_3 = c_2/2$ that results in:

$$\mathrm{SSIM}(x, y) = \frac{(2\mu_x\mu_y + c_1)(2\sigma_{xy} + c_2)}{(\mu_x^2 + \mu_y^2 + c_2)(\sigma_x^2 + \sigma_y^2 + c_2)}$$

We used SSIM to determine the quality of the reconstructed and generated objects in the main text.

### 3. Object detection from Single to multifrequency far-field radiation data

In the main text, we presented the results for five frequencies radiation data to determine the shape of object accurately. By including multifrequency data in the training process, we demonstrate how the results are significantly improved by providing training and prediction of neural networks with different frequencies.

#### 3.1 Object detection from Single frequency far-field radiation profile

We start our analysis with a single frequency($f_1 = 1$ kHz) radiation data and designed the neural network to solve the forward and inverse problem. The five-layer forward neural network is designed containing 4096 − 500 − 500 − 500− 400 − 400 −87 nodes. The LeakyReLu function is used for activation, and Adam optimization is used with a learning rate $10^{-4}$. The training and prediction results for the designed forward network are shown in Figs. S3(a) and S3(b), respectively. The relative prediction error is computed over the testing data with mean error 1.72% indicated by vertical red dashed line. Fig.S1(c) illustrates the comparison of computed radiation pattens for two given arbitary objects where the COMSOL results (solid blue curve) are in perfect agreement with ML predictions (red dashed curve).

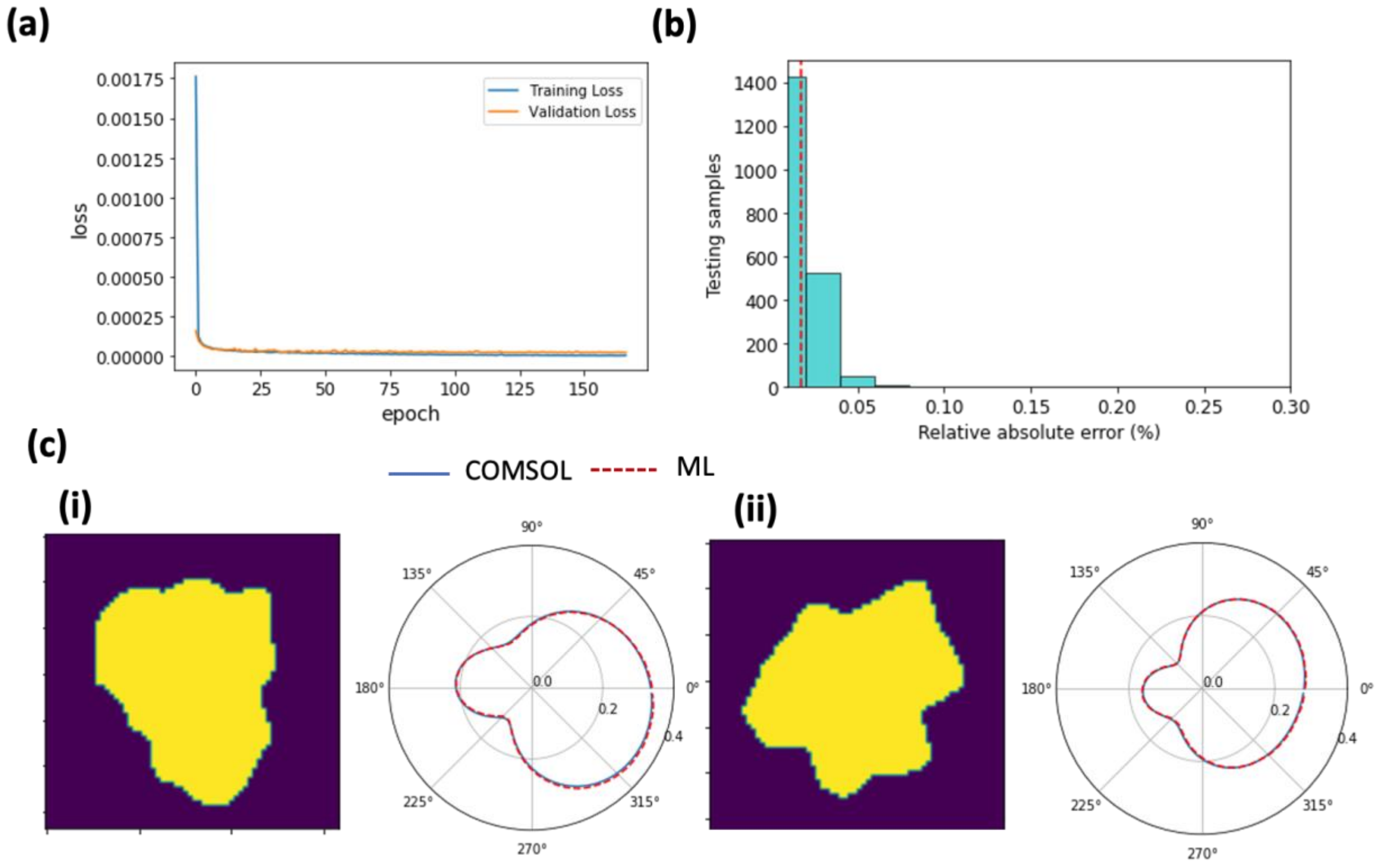

Fig.S3. Training and prediction performance of forward designed neural network for single frequency far-field radiation data. (a) Learning curve of optimized network. (b) Histogram of spectral prediction error. (c) Representative examples for far-field prediction from given arbitary objects. The solid blue and dotted red line represent the far-field response calculated from COMSOL and ML method, respectively.

Next, we proceed to solve the inverse problem of arbitary shape detection with single frequency far-field data. The inverse designed network contains 87− 800 − 500 − 500− 500 − 400 −100 nodes. The training and prediction results are presented in Fig.S4. The relative spectral error, binary cross-entropy error (BCE) and SSIM over generated objects for the testing data are computed with mean values of 0.7%, 0.271, and 0.70 respectively. Although the spectral error is very low but large BCN error in generated geometries as compared to target objects is expected due to presence of non-unique solution space, as depicted in Fig. S4(e).

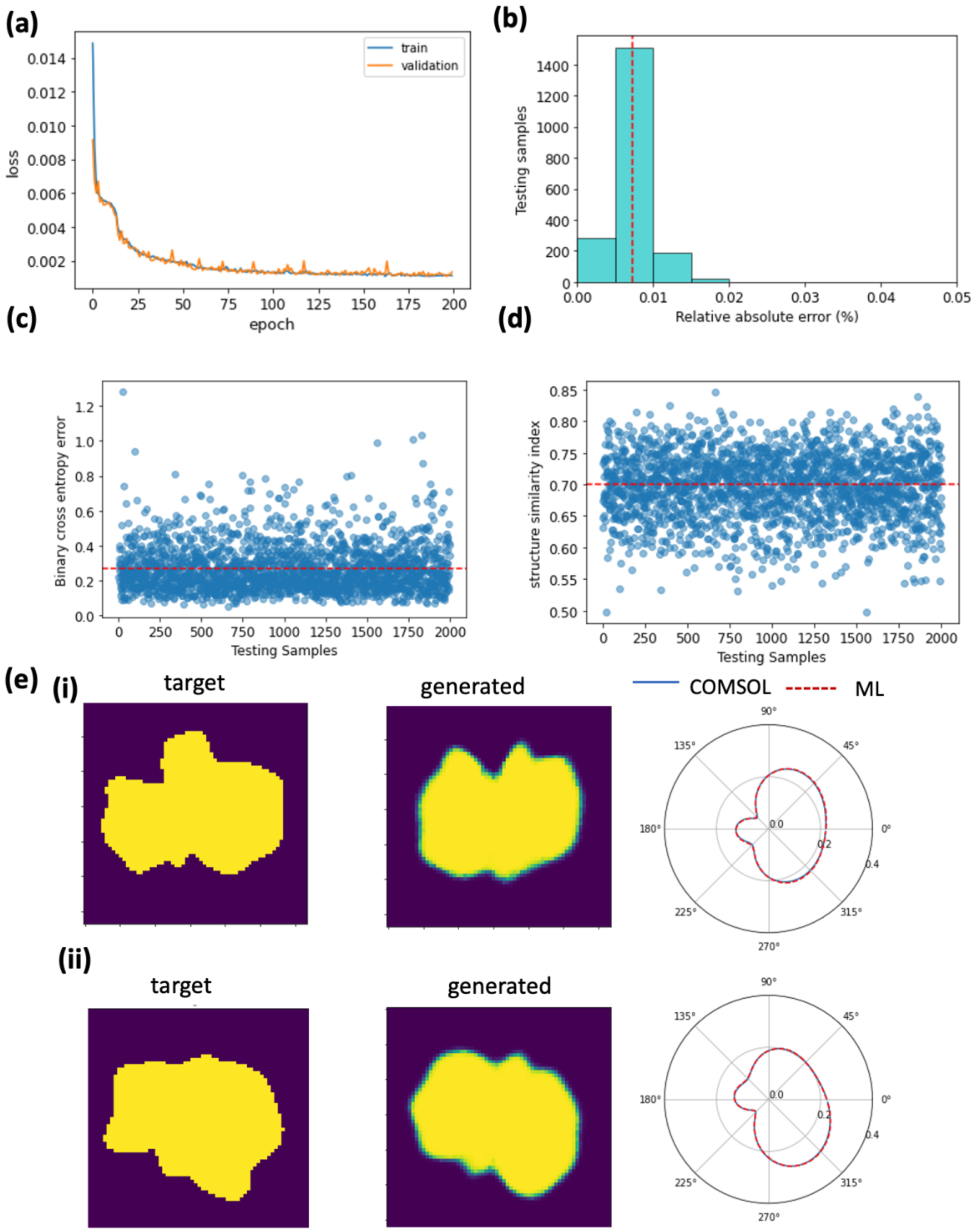

Fig. S4. Training and prediction performance of inverse designed neural network for single frequency far-field radiation data. a Learning curve of optimized network. (b) Histogram of spectral prediction error. (c) binary cross-entropy error for ML generated shapes. (d) distribution of SSIM for the ML generated geometries. (e) Representative examples for arbitary object prediction from given two-frequency far-field data. The predicted far-field response (dashed red) from the generated objects exactly matches with (solid blue) but the generated shapes show variations from the target objects due to non-unique solution space in inverse process.

### 3.2 Object detection from far-field radiation profile at two different frequencies

Next, we train the network with far-field radiation data at two different frequencies $f_1 = 1$ kHz and $f_2 = 1.5$ kHz. The forward and inverse architectures consist of $4096 - 500 - 500 - 500 - 400 - 174$ and $174 - 800 - 800 - 500 - 500 - 500 - 400 - 100$ nodes, respectively. The training and prediction results for forward and inverse network are shown in Figs. S5 and S6, respectively. On Fig.S5(b), the red line indicates the mean relative spectral error for trained forward network as 2.27%. In inverse network, the relative spectral error, binary cross-entropy error (BCE) and SSIM over generated objects for the testing data are computed with mean values of 1%, 0.131, and 0.75 respectively, as depicted in Fig. S8.

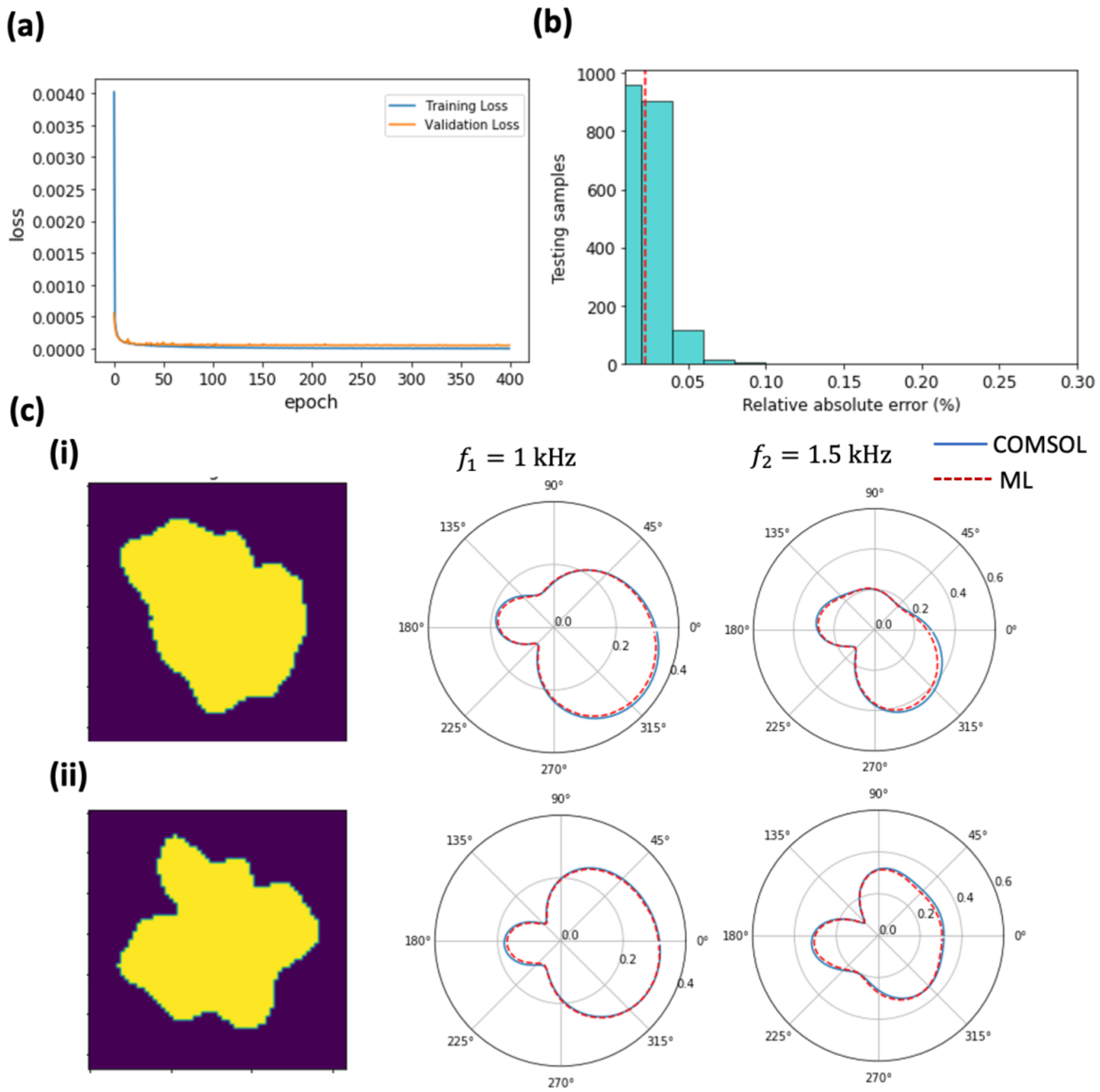

Fig.S5. Training and prediction performance of forward designed neural network for far-field radiation data at two different frequencies. (a) Learning curve of optimized network. (b) Histogram of spectral prediction error. (c) Representative examples for far-field prediction from given arbitary objects. The solid blue and dotted red line represent the far-field response calculated from COMSOL and ML method, respectively.

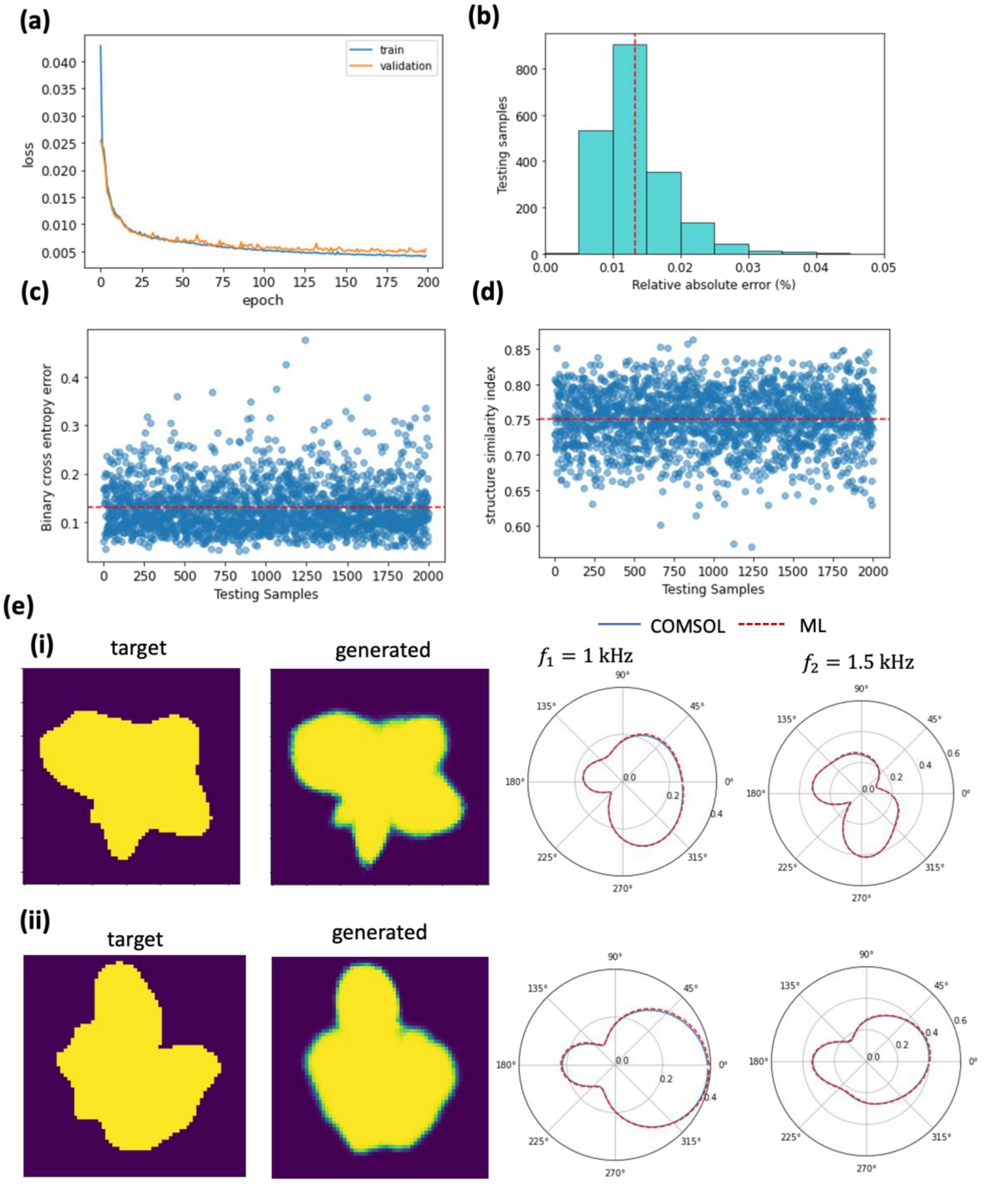

Fig. S6 Training and prediction performance of inverse designed neural network for two-frequency far-field radiation data. (a) Learning curve of optimized network. (b) Histogram of spectral prediction error. (c) binary cross-entropy error for ML generated shapes. d distribution of SSIM for the ML generated geometries. (e) Representative examples for arbitary object prediction from given two-frequency far-field data. The predicted far-field response (dashed red) from the generated objects exactly matches with (solid blue) but the generated shapes show variations from the target objects due to non-unique solution space in inverse process.

### 3.3 Object detection from far-field radiation profile at three different frequencies

Subsequently, we consider three different frequencies $f_1 = 1$ kHz $f_2 = 1.5$ kHz, and $f_3 = 2$ kHz. The results for the trained forward and inverse networks are shown in Figs. S7 and S8, respectively. The designed forward and inverse architectures consist of $4096 - 500 - 500 - 500 - 400 - 400 - 261$ and $174 - 800 - 800 - 500 - 500 - 500 - 400 - 100$ nodes, respectively. The red dashed line in Fig. S7(b) indicates the mean relative spectral error for the trained forward network as 2.7%. For the inverse case, relative spectral error, binary cross-entropy error (BCE), and SSIM over generated objects are computed with mean values of 1.96%, 0.0986, and 0.79, respectively [see Fig.S8].

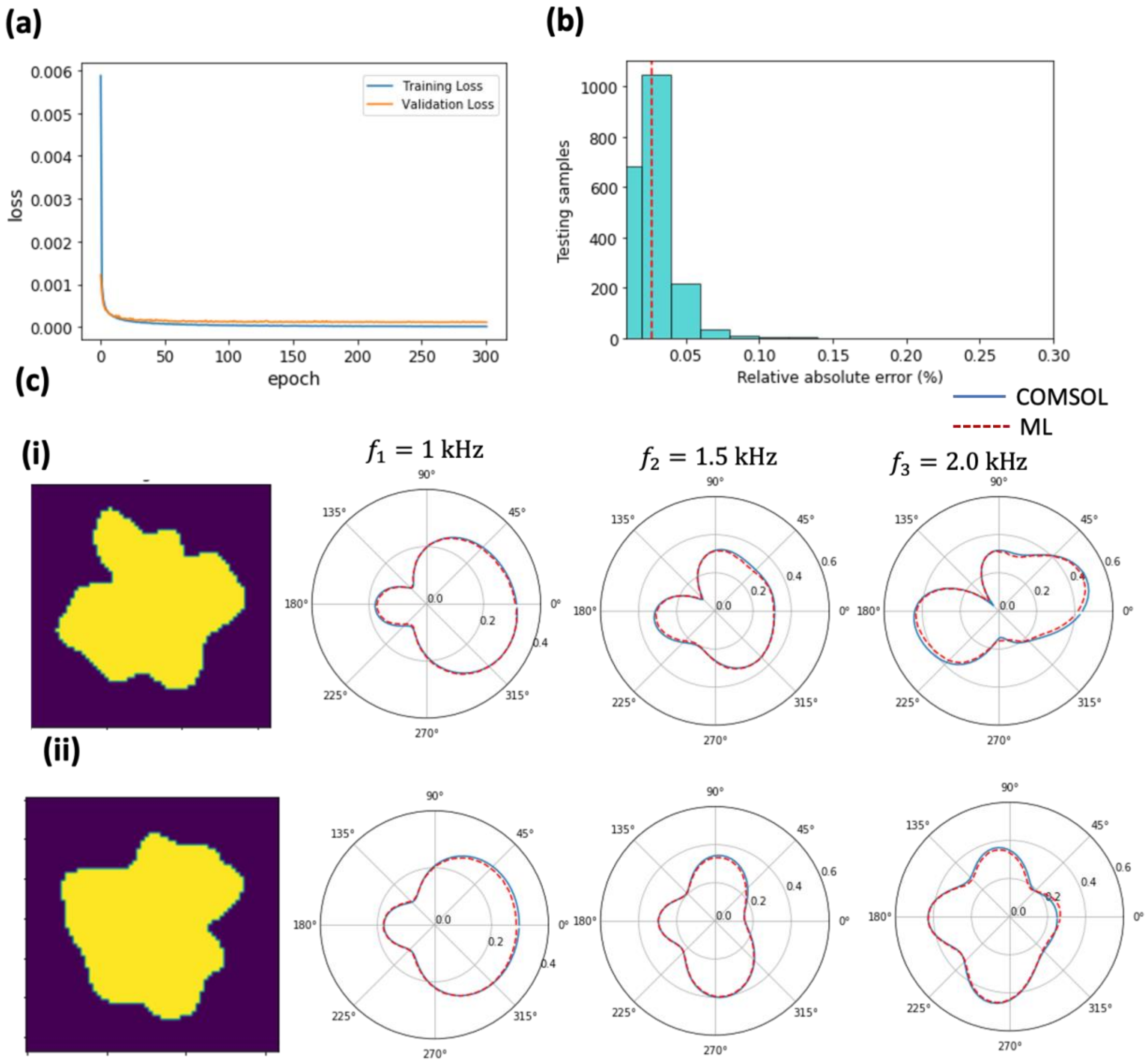

Fig. S7. Training and prediction performance of forward designed neural network for far-field radiation data at three different frequencies. (a) Learning curve of optimized network. (b) Histogram of spectral prediction error. (c) Representative examples for far-field prediction from given arbitary objects. The solid blue and dotted red line represent the far-field response calculated from COMSOL and ML method, respectively.

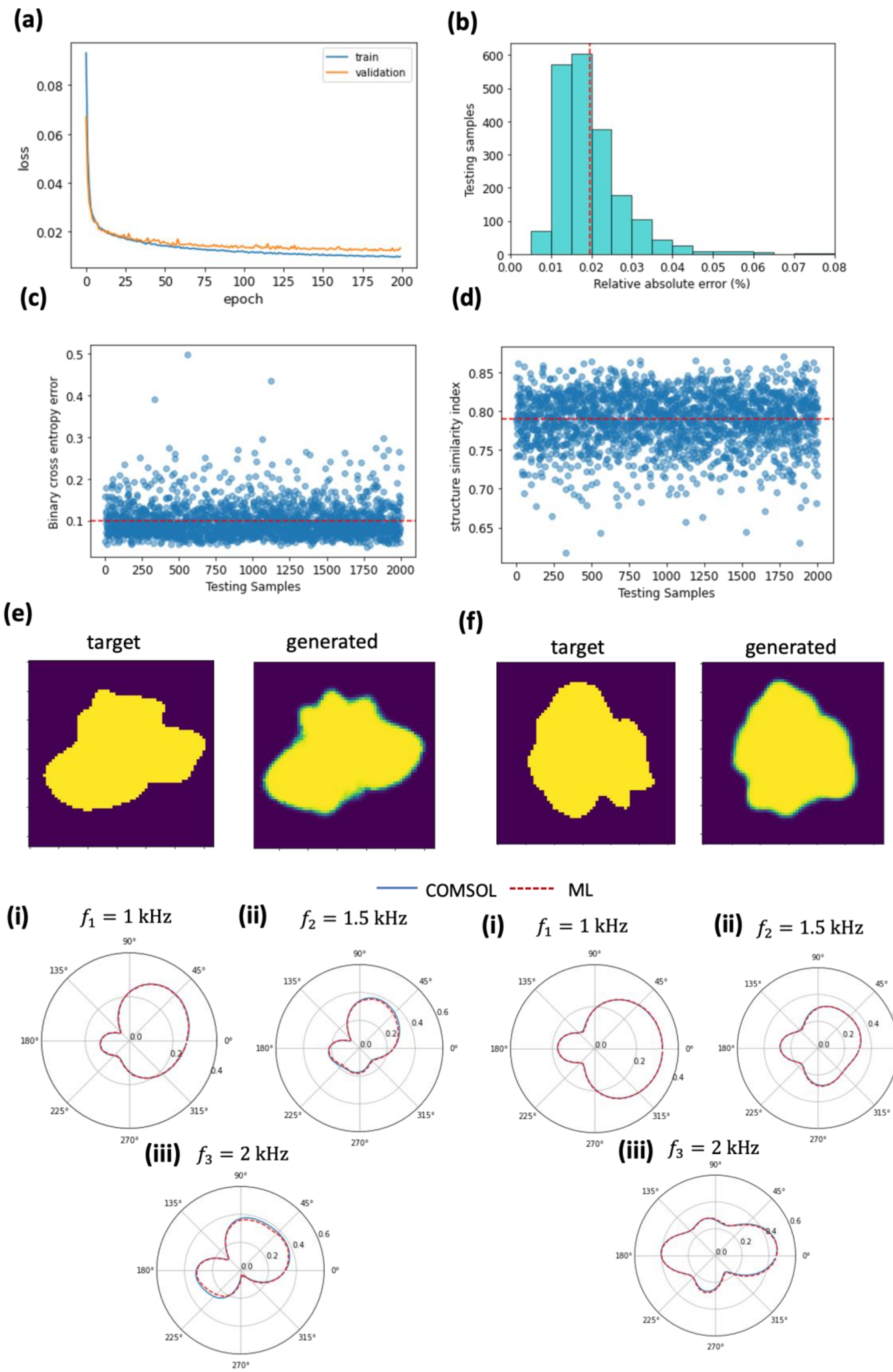

Fig. S8 Training and prediction performance of inverse designed neural network for far-field radiation data at three different frequencies. (a) Learning curve of optimized network. (b) Histogram of spectral prediction error. (c) binary cross-entropy error for ML generated shapes. d distribution of SSIM for the ML generated geometries. (e)

Representative examples for arbitary object prediction from given two-frequency far-field data. The predicted far-field response (dashed red) from the generated objects exactly matches with (solid blue) but the generated shapes show variations from the target objects due to non-unique solution space in inverse process. The results for arbitary shape detection are significantly improved due to addition of extra information in the training process as compared to Figs.S5 and S6.

### 3.4 Object detection from far-field radiation profile at four different frequencies

Lastly, we train the network with far-field radiation data at four different frequencies $f_1 =$ 1 kHz $f_2 = 1.5$ kHz, $f_3 = 2$ kHz and $f_4 = 2.5$ kHz and results for trained forward and inverse networks are presented in Figs. S9 and S10, respectively. The designed forward and inverse architectures consist of $4096 - 800 - 800 - 800 - 800 - 600 - 600 - 348$ and $348 - 800 - 800 - 500 - 500 - 500 - 400 - 100$ nodes, respectively. Fig.S(9) illustrates the distribution of relative spectral error over testing data with mean value 3.13%. Fig. S10 shows the inverse network's relative spectral error, binary cross-entropy error (BCE) and SSIM over the generated objects with mean values of 2.5%, 0.0916, and 0.83 respectively.

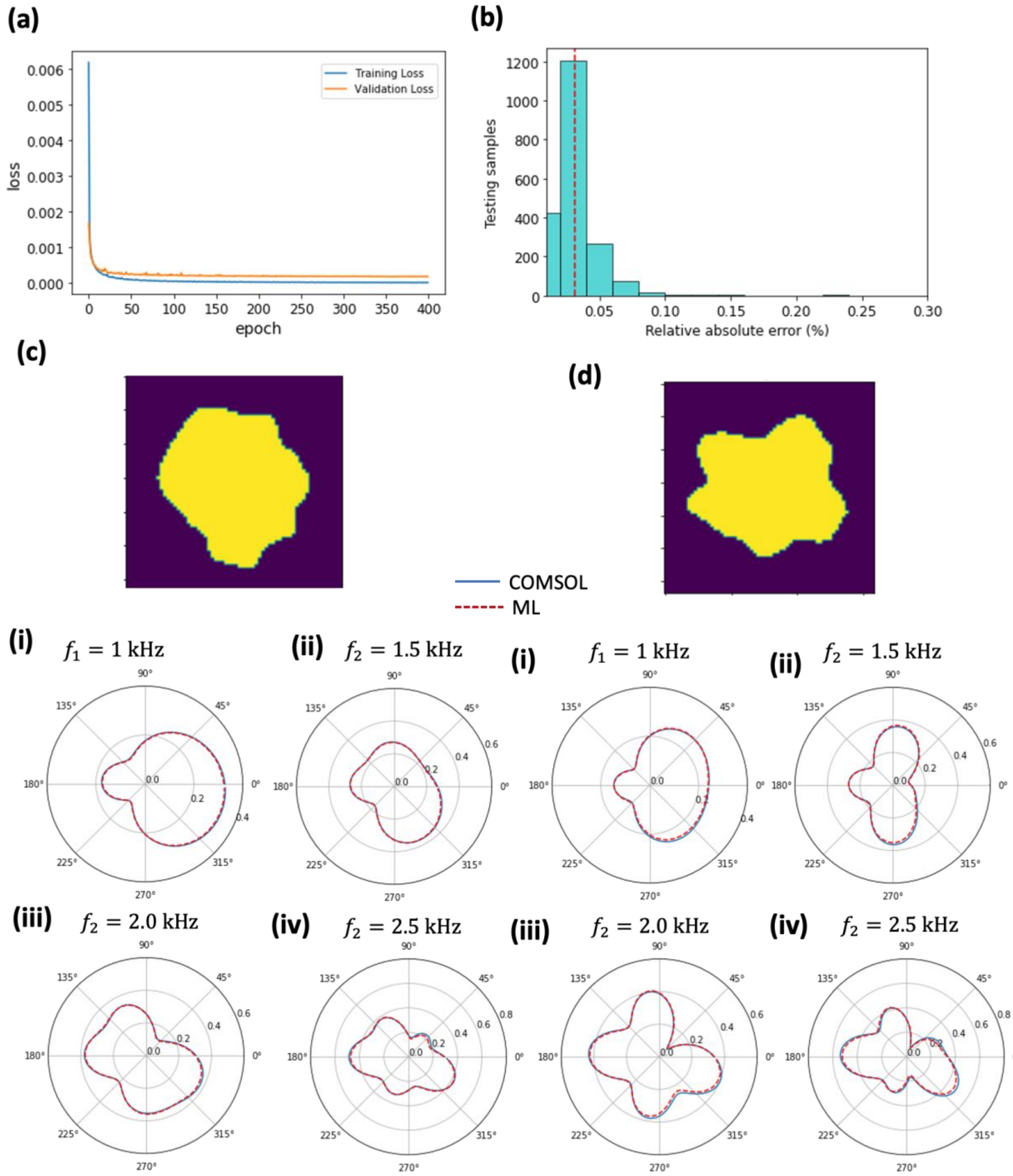

Fig. S9. Training and prediction performance of forward designed neural network for far-field radiation data at four different frequencies. (a) Learning curve of optimized network. (b) Histogram of spectral prediction error. (c) Representative examples for far-field prediction from given arbitary objects. The solid blue and dotted red line represent the far-field response calculated from COMSOL and ML method, respectively.

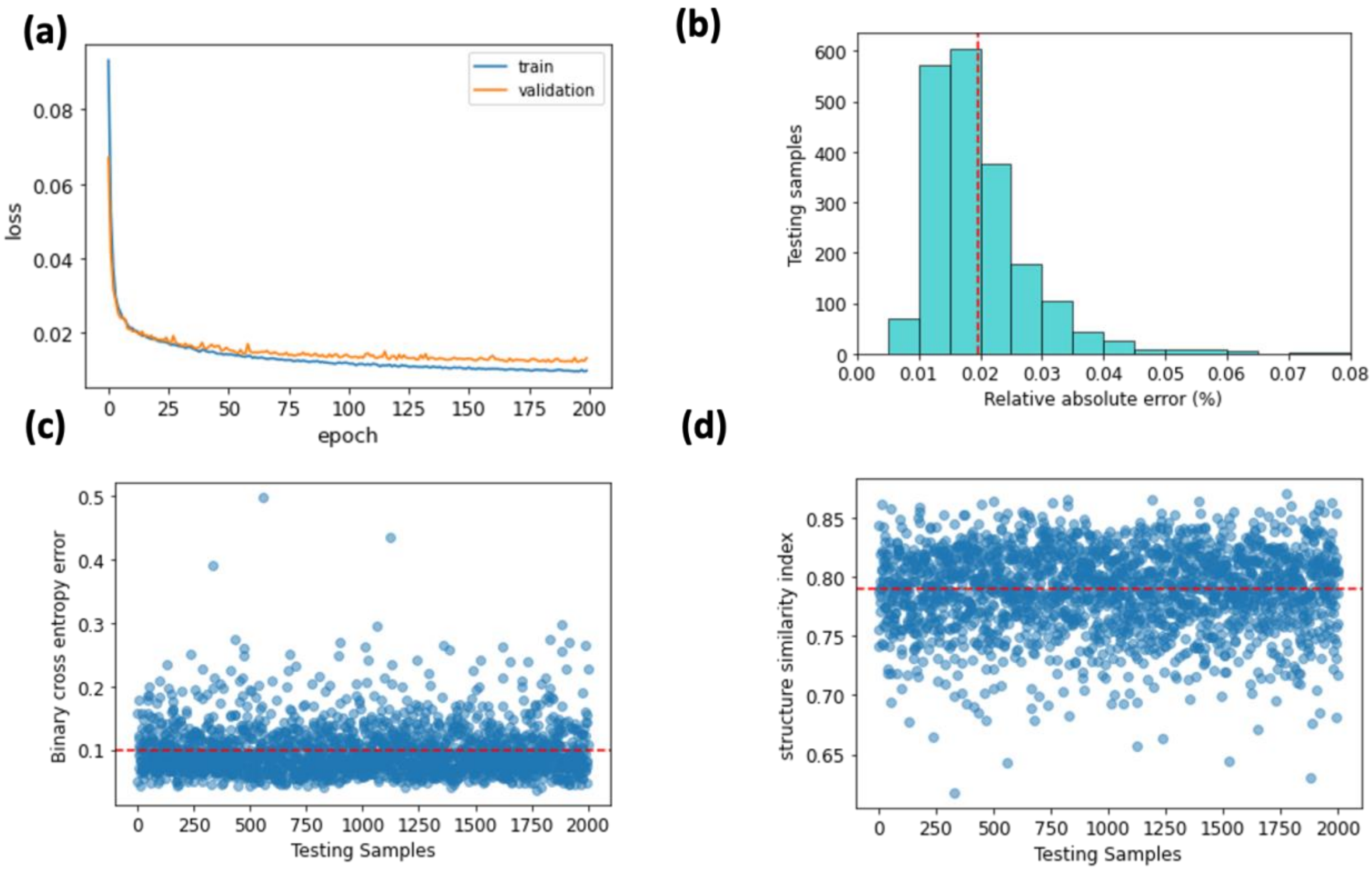

Fig. S10. Training and prediction performance of inverse designed neural network for far-field radiation data at four different frequencies. (a) Learning curve of optimized network. (b) Histogram of spectral prediction error. (c) binary cross-entropy error for ML generated shapes. d distribution of SSIM for the ML generated geometries.

## 4. Object detection from multifrequency half-plane far-field data

Here, we present the training behavior of designed neural network to predict the shape of arbitary object from multifrequency half-plane far-field data which is discussed in the main text. The designed forward and inverse architectures consist of $4096 - 1000 - 1000 - 800 - 800 - 600 - 600 - 328$ and $328 - 800 - 800 - 500 - 500 - 400 - 100$ nodes, respectively. In addition, some more examples are provided for arbitary shape detection from the given far-field patterns.

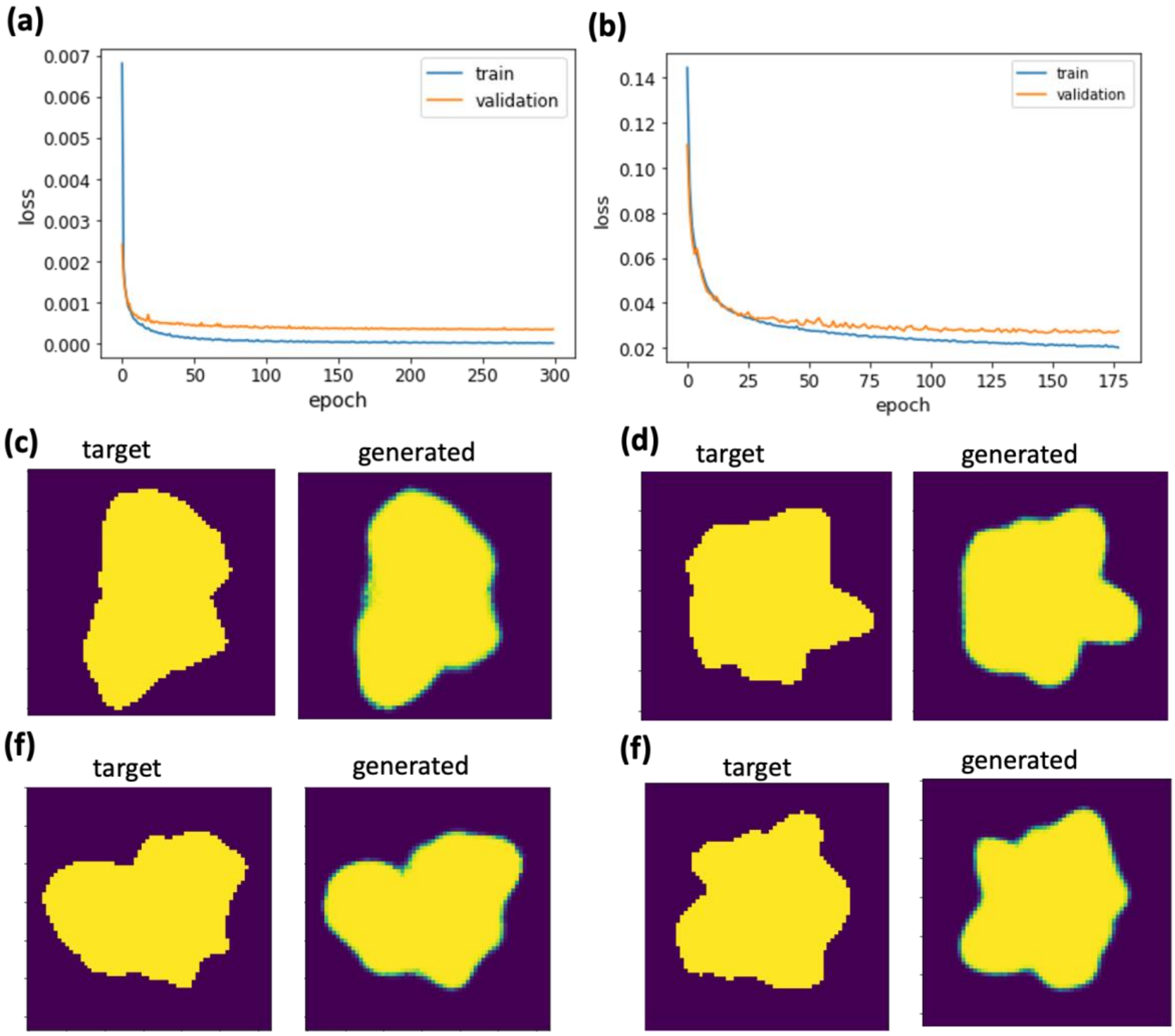

Fig. S11. Training performance of designed neural network for half- plane multifrequency far-field radiation data. (a) Learning curve of forward optimized network. (b) Learning curve of inverse optimized network. (c-f) target object and generated object from the inverse designed network.

## 5. More examples for reconstruction of arbitary Object from AAE

Here, we present some more examples to reconstruct the arbitary object from the designed AAE in the main text.

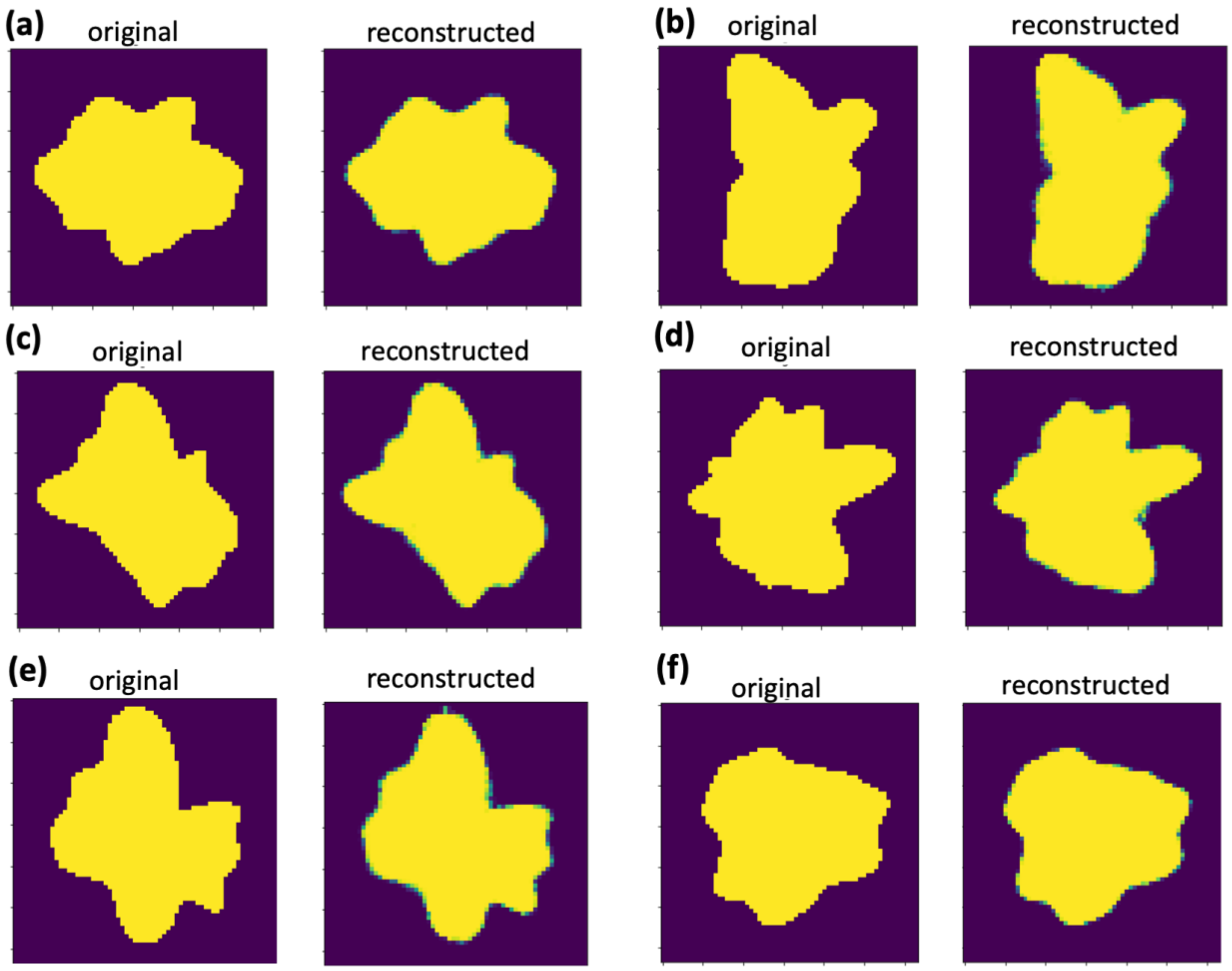

Fig. S12. Examples of Object reconstruction from trained adversarial autoencoder discussed in main text.